\documentclass[12pt]{JHEP3}
\pdfoutput=1

\usepackage{amsmath}
\usepackage{graphicx}
\usepackage{multirow}
\usepackage{epsfig}
\usepackage{rotating}
\usepackage{url}

\title{$Z'$ Bosons at Colliders: a Bayesian Viewpoint}
\author{Jens Erler$^a$, Paul Langacker$^b$, Shoaib Munir$^{a,c}$ and Eduardo Rojas$^a$
\\
$^a$Departamento de F\'isica Te\'orica, Instituto de F\'isica, \\ 
Universidad Nacional Aut\'onoma de M\'exico, 04510 M\'exico D.F., M\'exico \\
$^b$School of Natural Sciences, Institute for Advanced Study, \\ 
Einstein Drive, Princeton, NJ 08540, USA \\
$^c$Department of Physics, COMSATS Institute of Information Technology, \\
Defence Road, Lahore-54000, Pakistan \\ 
\\
\email{erler@fisica.unam.mx} \\
\email{pgl@ias.edu} \\
\email{smunir@fisica.unam.mx} \\
\email{eduardor@fisica.unam.mx}}

\newcommand{\be}{\begin{equation}}
\newcommand{\ee}{\end{equation}}
\newcommand{\bea}{\begin{eqnarray}}
\newcommand{\eea}{\end{eqnarray}}
\newcommand{\ba}{\begin{array}}
\newcommand{\ea}{\end{array}}
\newcommand{\nn}{\nonumber}
\newcommand{\ovl}{\overline} 
\newcommand{\ie}{\emph{i.e.,\ }}
\newcommand{\eg}{\emph{e.g.,\ }}
\newcommand{\sbx}{\scalebox}

\date{}
 
\abstract{We revisit the CDF data on di-muon production to impose constraints on a large class of $Z'$ bosons occurring in a variety of $E_6$ GUT based models. We analyze the dependence of these limits on various factors contributing to the production cross-section, showing that currently systematic and theoretical uncertainties play a relatively minor role. Driven by this observation, we emphasize the use of the Bayesian statistical method, which allows us to straightforwardly (i)~vary the gauge coupling strength, $g'$, of the underlying $U(1)'$; (ii)~include interference effects with the $Z'$ amplitude (which are especially important for large $g'$); (iii)~smoothly vary the $U(1)'$ charges; (iv)~combine these data with the electroweak precision constraints as well as with other observables obtained from colliders such as LEP 2 and the LHC; and (v)~find preferred regions in parameter space once an excess is seen. We adopt this method as a complementary approach for a couple of sample models and find limits on the $Z'$ mass, generally differing by only a few percent from the corresponding CDF ones when we follow their approach. Another general result is that the interference effects are quite relevant if one aims at discriminating between models. Finally, the Bayesian approach frees us of any {\em ad hoc\/} assumptions about the number of events needed to constitute a signal or exclusion limit for various actual and hypothetical reference energies and luminosities at the Tevatron and the LHC.}

\begin{document}

\section{Introduction}
\label{sec:intro}

The search for new physics beyond the Standard Model (SM) is one of the main objectives of the current and future collider experiments. 
A promising signature of such physics are neutral $Z'$ bosons which appear in numerous models extending the SM gauge symmetry
group by an additional $U(1)'$ factor (for reviews, see~\cite{Leike:1998wr,Rizzo:2006nw,Langacker:2008yv,Nath:2010zj}). 
This is not only interesting in its own right. 
In fact, many (or even most) theories, scenarios, and models beyond the SM have been supplemented by extra $U(1)'$ symmetries,
in order to cure or ease specific problems that have arisen there. 
Thus, by finding a $Z'$ and reconstructing the underlying $U(1)'$ charges, one may obtain clues regarding the underlying
physics and principles, like supersymmetry, large or warped extra dimensions, other strong dynamics, {\em etc.}

There is currently no experimental evidence for a $Z'$, 
not in the electroweak precision data (EWPD)~\cite{Erler:2009jh,delAguila:2010mx},
nor in the study of interference effects at LEP~2~\cite{Alcaraz:2006mx},
nor in searches for resonance production at the Tevatron~\cite{Jaffre:2009dg} or the LHC~\cite{CMS:2011wq,Collaboration:2011xp}.
It is therefore customary to set lower limits on the $Z'$ boson masses, $M_{Z'}$, for a number of models
and relative to a fixed value of the $U(1)'$ gauge coupling, $g'$.

However, one can extract more information from all the available experimental results than is reflected in a collection of mass limits.  
Indeed, the information is based on a variety of different channels and observables ({\em e.g.}, cross-sections and asymmetries) 
which can be used to disentangle the underlying model parameters and to diagnose the $Z'$.  
Moreover, precision constraints come from the $Z$-pole and related observables,
from low energy measurements, and from the flavor sector. 
Thus, one should find a framework allowing to discuss these very different sources simultaneously and transparently. 
Bayesian data analysis, being particularly suited for parameter estimation (as opposed to hypothesis and model
testing), proves very convenient to achieve this goal, and in this paper we take a first step in this direction. Specifically, we consider as an example the recent di-muon results by the CDF Collaboration~\cite{Aaltonen:2008ah}.



Working towards the above mentioned goal, we exploit here the implicit features of the Bayesian approach i) to study the effects of interference of the $Z'$ boson with the $\gamma$ and $Z$ bosons; and ii) to project exclusion limits for current, future, and hypothetical colliders and various luminosity reaches. For the latter, this approach avoids any {\em ad hoc\/} assumptions about how many observed or expected events would constitute an exclusion or a discovery. 

In our previous study~\cite{Erler:2009jh} of $Z'$ bosons we analyzed the most recent EWPD, which --- as usual ---
was based on least-$\chi^2$ fits, \ie the likelihood is given by $ {\cal L}_{\rm EWPD} = \exp(-\chi^2/2)$.
Further factors ${\cal L}_i$ entering the total (posterior) probability density need to be constructed for each data set,
\be
\label{eq:posterior}
   p_{\rm posterior}(M_{Z'}, g', \theta_{ZZ'}, ... ) = 
   p_{\rm prior}(M_{Z'}, g', \theta_{ZZ'}, ... ) \times {\cal L}_{\rm EWPD}  \times \prod_i {\cal L}_i,
\ee
where $\theta_{ZZ'}$ is the mixing angle between the $Z'$ and the ordinary $Z$ boson 
and $p_{\rm prior}$ is a non-informative prior density. 
We usually take it to be flat for variables defined over ${\cal R}^n$ having in mind an infinitely wide multivariant Gaussian distribution.  
In other cases a parameter transformation may be in order.
The dots refer to further $Z'$ parameters such as those characterizing the $U(1)'$ charges. 
We believe that computing the full $p_{\rm posterior}$ is a (long-term) task worth its effort to make full use of all experimental
(and theoretical) information.  
This is particularly obvious when a signal is seen in one or several places, and one wants to narrow down the space of possible
underlying $U(1)'$ symmetries, as mentioned above. 
Notice that $p_{\rm posterior}$ can easily be updated by regarding it as a (now informative) prior density and multiplying it with
a new factor of ${\cal L}_i$ (we are assuming here that there are no experimental or theoretical correlations between various factors
${\cal L}_i$ so that the factorization property holds).

Since among the final goals of our approach is the ability to discriminate among models (charges) we review in Section~\ref{sec:models}
a large and popular class of $Z'$ models based on the $E_6$ gauge group and try to put these on an equal footing. 
$Z'$ models with the same charges (at least as far as the SM fermions are concerned) 
also arise from a bottom-up approach~\cite{Erler:2000wu} when demanding 
the cancellation of gauge and mixed gauge-gravitational anomalies in supersymmetric extensions of the SM 
together with a set of fairly general requirements such as allowing the SM Yukawa couplings,
gauge coupling unification, 
a solution~\cite{Suematsu:1994qm,Cvetic:1995rj} to the $\mu$-problem~\cite{Kim:1983dt}, 
the absence of dimension~4 proton decay as well as fractional electric charges,
and chirality (to protect all fields from acquiring very large masses). 

In Section~\ref{sec:formalism} we lay out the theoretical framework for $Z'$ hadro-production, where 
some technical details relevant to the calculation of cross-sections are given for completeness in Appendix~\ref{app:formalism}. 
As a check, we reproduce the CDF limits following their approach and 
extend the limits to models not considered in their original analysis. 
Moreover, we project $M_{Z'}$ limits for various integrated luminosities and center-of-mass (CM) energies for the LHC. 

Finally, in Section~\ref{sec:Bayes} we discuss the proposed Bayesian statistical method.
We derive mass limits and compare them with those in Section~\ref{sec:formalism}.
We then compute exclusion contours for some illustrative models, emphasizing the role of interference between the $Z'$ and
SM amplitudes, especially for large $g'$. We conclude in Section~\ref{sec:conclusions}.

\section{The $E_6$ model class}
\label{sec:models}

As mentioned in the introduction, a very large class of $U(1)'$ symmetries underlying the $Z'$ bosons are subgroups of $E_6$,
and can be written in the form,
\be
\label{eq:e6models}
Z' = \cos\alpha \cos\beta\, Z_\chi + \sin\alpha \cos\beta\,  Z_Y + \sin\beta\,  Z_\psi 
    = {c_1\, Z_R + \sqrt{3}\, (c_2\, Z_{R_1} + c_3\, Z_{L_1}) \over \sqrt{c_1^2 + 3\, (c_2^2 + c_3^2)}}.
\ee
Here $-\pi/2 < \beta \leq \pi/2$ is the mixing angle between the $U(1)_\chi$ and $U(1)_\psi$ maximal subgroups 
defined by~\cite{Robinett:1982tq} $SO(10) \to SU(5) \times U(1)_\chi$ and $E_6 \to SO(10) \times U(1)_\psi$, respectively,
and $-\pi/2 < \alpha \leq \pi/2$ is non-vanishing when there is a mixing term~\cite{Holdom:1985ag} 
between the hypercharge, $Y$, and $U(1)'$ field strengths $\propto F_Y^{\mu\nu} F'_{\mu\nu}$,
and this kinetic mixing term has been undone by field redefinitions. 
The $U(1)_Y$, $U(1)_\chi$, and $U(1)_\psi$ groups are mutually orthogonal in the sense that 
$\rm{Tr}(Q_i Q_j) = 0$ when the trace is taken over a complete representation of $E_6$. 
The second form appearing in Eq.~(\ref{eq:e6models}) uses a different orthogonal basis,
$U(1)_R$, $U(1)_{R_1}$, and $U(1)_{L_1}$, which are the maximal subgroups~\cite{Robinett:1982tq} 
defined by $SU(3)_{L,R} \to SU(2)_{L,R} \times U(1)_{L_1,R_1}$ and $SU(2)_R \to U(1)_R$, referring here 
to the trinification subgroup~\cite{Achiman:1978vg} of $E_6 \to SU(3)_C \times SU(3)_L \times SU(3)_R$.

\TABLE[t]{\label{tab:e6charges}
\begin{tabular}{|ccrr|ccrr|}
\hline
$l \equiv \left( \ba{c} \nu \\ e^- \ea \right)$ & & $-2 c_2$ & $-c_3$ &
$\ba{c} \bar{\nu} \\ e^+ \ea$ & $\ba{r} -c_1 \\ +c_1 \ea$ & $\ba{r} +c_2 \\ +c_2 \ea$ & $\ba{r} +2 c_3 \\ +2 c_3 \ea$ \\
\hline
$q \equiv \left( \ba{c} \phantom{l} u \phantom{l} \\ \phantom{l} d \phantom{l} \ea \right)$ & & & $+c_3$ &
$\ba{c} \bar{u} \\ \bar{d} \ea$ & $\ba{r} -c_1 \\ +c_1 \ea$ & $\ba{r} -c_2 \\ -c_2 \ea$ & \\
\hline\hline
$L \equiv \left( \ba{c} N \\ E^- \ea \right)$ & $-c_1$ & $+c_2$ & $-c_3$ &
$\ba{c} D \\ \ovl{D} \ea$ & & $\ba{r} \\ +2 c_2 \ea$ & $\ba{r} -2 c_3 \\ {} \ea$ \\
\hline
$\ovl{L} \equiv \left( \ba{c} E^+ \\ \ovl{N} \ea \right)$ & $+c_1$ & $+c_2$ & $-c_3$& 
$S$ & & $\ba{r} -2 c_2 \ea$ & $\ba{r} +2 c_3 \ea$ \\
\hline
\end{tabular}
\caption{Charge assignment for the left-handed multiplets contained in a $\bf{27}$ dimensional representation of $E_6$.
The upper part of the table corresponds to the $\bf{16}$ dimensional representation of $SO(10)$, while the lower part shows the 
$\bf{10}$ (with an extra anti-quark weak singlet, $\ovl{D}$, of electric charge $-1/3$ and an additional weak doublet, $L$,
as well as their SM-mirror partners) and the~$\bf{1}$ (a SM singlet, $S$).  
This represents one fermion generation, and we assume family universality throughout.
The correct normalization (\ie the one which is directly comparable to the usual normalization of the gauge couplings of $SU(3)_C$
and $SU(2)_L$ of the SM) of these charges is obtained upon division by $2 \sqrt{c_1^2 + 3\, (c_2^2 + c_3^2)}$.}}

The $U(1)'$ charges of the particles appearing in the fundamental representation of $E_6$ are shown in 
Table~\ref{tab:e6charges} in terms of the parameters $c_1$, $c_2$ and $c_3$, satisfying,
\be
   \tan\alpha = {c_1 + c_2 + c_3 \over \sqrt{2\over 3}\, c_1 - \sqrt{3\over 2}\, (c_2 + c_3)}, \hspace{24pt}
   \tan\beta = {{\rm sgn} [{2 \over 3}\, c_1 - (c_2 + c_3)] \over \sqrt{{2\over 3}\, c_1^2 + (c_2 + c_3)^2}}  (c_3 - c_2).
\ee
The values of $\alpha$, $\beta$, and the $c_i$ for some specific models are given in Table~\ref{tab:models}.
We also display the charges for the models listed in Table~\ref{tab:models} more explicitly in Table~\ref{tab:SMcharges}.
The general classification of all models with integer charges, as well as models arising from breaking chains
involving maximal subgroups is the subject of Ref.~\cite{ER:2011}. 

\TABLE[t]{\label{tab:models}
\begin{tabular}{|l|rrr|rr|}
\hline
$\phantom{-} Z^\prime$ & $c_1$ & $c_2$ & $c_3$ & $\tan \alpha$ & $\tan \beta$ \\
\hline
$\phantom{-} Z_R$~\cite{Robinett:1982tq}                & $+1$   & $0$    & $0$  & $+\sqrt{3/2}$  & $0$ \\
$\phantom{-} Z_{\not d}$                                       & $-1/2$ & $-1/2$ & $0$  & $-\sqrt{24}$   & $+\sqrt{3/5}$ \\
$                  -  Z_I$~\cite{Robinett:1982tq}       & $+1/2$ & $-1/2$ & $0$  & $0$            & $+\sqrt{3/5}$ \\
$                  -  Z_{L_1}$~\cite{Robinett:1982tq}   & $0$    & $0$    & $-1$ & $-\sqrt{2/3}$  & $-1$ \\
$                  -  Z_{R_1}$~\cite{Robinett:1982tq}   & $0$    & $-1$   & $0$  & $-\sqrt{2/3}$  & $+1$ \\
$\phantom{-} Z_{\not p}$                                & $+3/2$ & $-1/2$ & $0$  & $+\sqrt{8/27}$ & $+1/\sqrt{7}$ \\
$                  -  Z_{\not n}$                       & $+3/2$ & $+1/2$ & $0$  & $+\sqrt{32/3}$ & $-1/\sqrt{7}$ \\
$                  -  Z_{B-L}$~\cite{Pati:1974yy}       & $0$    & $-1$   & $-1$ & $-\sqrt{2/3}$  & $0$ \\
$\phantom{-} Z_{ALR}$~\cite{Ma:1986we}                  & $+3/2$ & $-1/2$ & $+1$ & $+\sqrt{32/3}$& $+3/\sqrt{7}$ \\
$                  -  Z_{\not L}$~\cite{Babu:1996vt}    & $+3/2$ & $+1/2$ & $-1$ & $+\sqrt{8/27}$ & $-3/\sqrt{7}$ \\
$\phantom{-} Z_\psi$~\cite{Robinett:1982tq}             & $0$    & $-1$   & $+1$ & ---            & $+\infty$ \\
$\phantom{-} Z_\chi$~\cite{Robinett:1982tq}             & $+2$   & $-1$   & $-1$ & $0$            & $0$ \\
$\phantom{-} Z_N$~\cite{Ma:1995xk,King:2005jy}          & $+1/2$ & $-3/2$ & $+1$ & $0$            & $+\sqrt{15}$ \\
$\phantom{-} Z_\eta$~\cite{Witten:1985xc}               & $+3/2$ & $+1/2$   & $-2$ & $0$            & $-\sqrt{5/3}$ \\
$\phantom{-} Z_Y$~\cite{Glashow:1961tr,Weinberg:1967tq} & $+3$   & $+1$   & $+1$ & $+\infty$      & $0$ \\
$\phantom{-} Z_S$~\cite{Erler:2002pr,Kang:2004pp}       & $+9/2$ & $-7/2$ & $-1$ & $0$            & $+\sqrt{5/27}$ \\
\hline 
\end{tabular}
\caption{The values of the $c_i$ and $(\alpha,\beta)$ parameters for various $E_6$ motivated $Z'$ bosons, 
most of them appearing in the literature (if referenced).
The $Z_{\not p}$ and the $Z_{\not n}$ are bosons which do not couple --- at vanishing momentum transfer and at the tree level ---
to protons and neutrons, respectively.
Similarly, the $Z_{\not L}$, $Z_I$, and $Z_{\not d}$ bosons are blind, respectively, to SM leptons, up-type quarks, and down-type quarks.
The $Z_{B-L}$ couples purely vector-like while the $Z_\psi$ has only axial-vector couplings to the ordinary fermions. 
The overall sign of the $c_i$ for each model is not physical, and can be absorbed into the definition of the $U(1)'$ gauge field 
when allowing both signs for the mixing angle, $\theta_{ZZ'}$.
However, the sign convention of the charges becomes significant once a sign convention for $\theta_{ZZ'}$ has been adopted. 
Where applicable we follow the sign conventions of Refs.~\cite{Langacker:2008yv,Erler:2009jh}
and otherwise the more systematic sign convention detailed in Ref.~\cite{ER:2011}.}}

There are also classes of models described by one continuous parameter.
For example, one can restrict oneself to $U(1)'$ subgroups of $SO(10)$, 
\ie those perpendicular to the $U(1)_\psi$ and therefore with $c_2 = c_3 \Longleftrightarrow \beta = 0$.
These models are equivalent (up to non-chiral sets of fermions) to the ones described by the real parameter $x$ 
and denoted by $U(1)_{q+xu}$ in Ref.~\cite{Carena:2004xs},
\ie with charges defined by $Q'_{\bar{u}} = -x Q'_q = x Q'_l/3$, when one identifies,
\be
   \tan\alpha = \sqrt{3 \over 2}\, {x+1 \over x-4}.
\ee
This class~\cite{Appelquist:2002mw} contains the $Z'$ models based on left-right symmetry, $Z_{LR}$,
which can be seen from the breaking, $SO(10) \to SU(3)_C \times SU(2)_L \times SU(2)_R \times U(1)_{B-L}$. 
Incidentally, Table~\ref{tab:models} shows that $U(1)_{B-L}$ is the diagonal combination of $U(1)_{L_1}$ and $U(1)_{R_1}$
(and indeed was initially dubbed $U(1)_{L+R}$~\cite{Pati:1974yy}) and has manifestly left-right symmetric charges.
However, left-right symmetry is broken in the SM.
In the fully $SO(10)$ symmetric case, 
\ie when all gauge couplings are equal, $g \equiv g_L = g_R = \tilde{g}_{B-L} =  \tilde{g}_Y$
(the tilde denotes $SO(10)$ normalization which we use here to simplify the discussion), the 
\be\label{eq:LR}
   Z_{LR} = \cos\theta_{LR}\, (-Z_{B-L}) + \sin\theta_{LR}\, Z_R
\ee
must be orthogonal to the $Z_Y$ since these are obtained by an $SO(2)$ rotation of $Z_{B-L} \perp Z_R$.
This yields the $Z_\chi$ and $\tan\theta_{LR} = \sqrt{2/3} \approx 39^\circ$.
At lower energies, renormalization group (RG) effects will generally split the gauge couplings.
One then has~\cite{Langacker:2008yv,Robinett:1981yz}
\be
   \tan\theta_{LR} = \sqrt{2 \over 3}\, {g_R \over \tilde{g}_{B-L}}
                              = \sqrt{{5 \over 3}\, {g_R^2 \over \tilde{g}_Y^2} - 1}
                              = \sqrt{{g_R^2 \over g_L^2} \cot^2\theta_W- 1},
\ee
where the second step uses the relation~\cite{Appelquist:2002mw}
$5\, \tilde{g}_Y^{-2} = 2\, \tilde{g}_{B-L}^{-2} + 3\, g_R^{-2} $,
and where the weak mixing angle $\theta_W \equiv \arctan (g_Y/g_L)$ appears in the last.
Thus, assuming manifest left-right symmetry, $g_L = g_R$, one finds $\theta_{LR} \approx 57^\circ$.
Formally one has the entire range, $0 \leq \theta_{LR} < 90^\circ$, 
but realistic breaking patterns~\cite{Robinett:1981yz} suggest $35^\circ \lesssim \theta_{LR} \lesssim 42^\circ$.
Finally, this range can be extended to include all $SO(10)$ models by identifying $\theta_{LR} = \alpha + \arctan\sqrt{2/3}$.

\TABLE[t]{\label{tab:SMcharges}
\begin{tabular}{|l||rrr|rr|r||rr|rr||r|}
\hline
$\phantom{2}\, Q'$ & $q$ & $\bar{u}$ & $e^+$ & $\bar{d}$ & $l$ & $\bar{\nu}$ & $\ovl{D}$ & $L$ & $D$ & $\ovl{L}$ & $S$ \\
\hline
$2\, Q_R$                 & $0$  & $-1$ & $+1$ & $+1$ & $0$  & $-1$ & $0$  & $-1$ & $0$ & $1$  & $0$  \\
$2\, Q_{\not d}$          & $0$  & $+1$ & $-1$ & $0$  & $+1$ & $0$  & $-1$ & $0$  & $0$ & $-1$ & $1$  \\
$2\, Q_I$                 & $0$  & $0$  & $0$  & $-1$ & $-1$ & $+1$ & $1$  &$1$   &$0$  &$0$   &$-1$  \\
$2 \sqrt{3}\, Q_{L_1}$    & $+1$ & $0$  & $+2$ & $0$  & $-1$ & $+2$ & $0$  &$-1$  &$-2$ &$-1$  &$2$   \\
$2 \sqrt{3}\, Q_{R_1}$    & $0$  & $-1$ & $+1$ & $-1$ & $-2$ & $+1$ & $2$  &$1$   &$0$  &$1$   &$-2$  \\
$2 \sqrt{3}\, Q_{\not p}$ & $0$  & $-1$ & $+1$ & $+2$ & $+1$ & $-2$ & $-1$ & $-2$ &$0$  &$1$   &$1$   \\
$2 \sqrt{3}\, Q_{\not n}$ & $0$  & $+2$ & $-2$ & $-1$ & $+1$ & $+1$ & $-1$ & $1$  &$0$  &$-2$  &$1$   \\
$2 \sqrt{6}\, Q_{B-L}$    & $+1$ & $-1$ & $+3$ & $-1$ & $-3$ & $+3$ & $2$  &$0$   &$-2$ &$0$   &$0$   \\
$2 \sqrt{6}\, Q_{ALR}$    & $+1$ & $-1$ & $+3$ & $+2$ & $0$  & $0$  & $-1$ &$-3$  &$-2$ &$0$   &$3$   \\
$2 \sqrt{6}\, Q_{\not L}$ & $+1$ & $+2$ & $0$  & $-1$ & $0$  & $+3$ & $-1$ &$0$   &$-2$ &$-3$  &$3$   \\
$2 \sqrt{6}\, Q_\psi$     & $+1$ & $+1$ & $+1$ & $+1$ & $+1$ & $+1$ & $-2$ &$-2$  &$-2$ &$-2$  &$4$   \\
$2 \sqrt{10}\, Q_\chi$    & $-1$ & $-1$ & $-1$ & $+3$ & $+3$ & $-5$ & $-2$ &$-2$  &$2$  &$2$   &$0$   \\
$2 \sqrt{10}\, Q_N$       & $+1$ & $+1$ & $+1$ & $+2$ & $+2$ & $0$  & $-3$ &$-3$  &$-2$ &$-2$  &$5$   \\
$2 \sqrt{15}\, Q_\eta$    & $+2$ & $+2$ & $+2$ & $-1$ & $-1$ & $+5$ & $-1$ &$-1$  &$-4$ &$-4$  &$5$   \\
$2 \sqrt{15}\, Q_Y$       & $+1$ & $-4$ & $+6$ & $+2$ & $-3$ & $0$  & $2$  &$-3$  &$-2$ &$3$   &$0$   \\
$4 \sqrt{15}\, Q_S$       & $-1$ & $-1$ & $-1$ & $+8$ & $+8$ & $-10$& $-7$ &$-7$  &$2$  &$2$   &$5$   \\
\hline
\end{tabular}
\caption{Explicit $U(1)'$ charges for the models defined in Table~\ref{tab:models}.
The normalization imposed by $E_6$ symmetry is also shown, and we will use this normalization throughout 
in order to avoid spurious factors in the discussion in the text.
Columns within single lines fill out $SU(5)$ multiplets, while double lines enclose full $SO(10)$ multiplets.}}

Similarly, there is a class of models perpendicular to the $U(1)_Y$ and therefore with 
$c_1 + c_2 + c_3 = 0 \Longleftrightarrow \alpha = 0$. Under the identification,
 \be
   \tan\beta = \sqrt{3\over 5}\, {x+3 \over x-1},
\ee
these models are equivalent to those denoted by $U(1)_{10+x\ovl{5}}$~\cite{Carena:2004xs}, 
\ie with charges related by $Q'_{\bar{d}} = x Q'_{\bar{u}} = x Q'_q$.
Finally, Ref.~\cite{Carena:2004xs} discussed another one-parameter subset of models, $U(1)_{d-xu}$,
which can be obtained by demanding $Q'_q = 0 \Longleftrightarrow c_3 = 0$ and $Q'_{\bar{u}} = -x Q'_{\bar{d}}$,
and by identifying,
\be
   \tan\alpha = - {2 \sqrt{6}\, x \over x-5},  \hspace{24pt}
   \tan\beta = {\sqrt{3}\, (x-1)\, {\rm sgn} (x-5) \over \sqrt{5 x^2 - 2 x + 5}}.
\ee
Of course, any other one-parameter subset of models may be considered.
All models are guaranteed to be free of anomalies due to the absence of an independent cubic Casimir invariant from $E_6$.
On the other hand, the $U(1)_{B-xL}$ model class~\cite{Carena:2004xs} is not contained in $E_6$ except for $x=1$,
and a different anomaly-free completion of the model (\eg involving charged lepton singlets~\cite{Carena:2004xs}) is needed.
A similar remark applies to the models in Ref.~\cite{Erler:2000wu} which predict $Z'$ charges of the SM fermions as in $E_6$ 
but distinct charges of exotics. {\em E.g.\/}, the $Z_{\tilde{\psi}}$ model~\cite{Erler:2000wu} couples like the $Z_\psi$ to the SM fermions
as well as to the exotic charged leptons, but the $D$ and $\ovl{D}$ charges are multiplied by a factor of 3/2 and further SM singlets
must be added (see, \eg Table III in Ref.~\cite{Langacker:2008yv} for details and generalizations).

Finally, we will consider two models corresponding to maximal constructive ($Z^+_{u-int}$) and destructive ($Z^-_{u-int}$) interference 
of the $Z'$ amplitude for $u$ quarks (dominating the Drell-Yan production process at large momentum transfer) 
with those of the $\gamma$ and the ordinary $Z$ boson (see Section~\ref{sec:Bayes}).

\section{Direct searches at hadron colliders}
\label{sec:formalism}

Due to the large QCD background at the Tevatron, the decay into a lepton pair is the preferred discovery channel for a $Z'$ since leptons 
are relatively easy to identify and their energies and momenta can be measured more precisely than those of hadrons, although
$b$ quarks~\cite{Abe:1996yya,Han:2004zh}, $t$ quarks~\cite{Han:2004zh,Aaltonen:2007dz}
and di-jets~\cite{Abe:1995jz,Abe:1996mj,Rizzo:1993ie} can also be detected. 
Among leptons, the background for a $\tau$ pair is harder to manage~\cite{Anderson:1992jz} compared to the $\mu^+\mu^-$ and 
$e^+e^-$ channels, of which the former is preferable still~\cite{Abe:1997fd}. 

\FIGURE[t]{\label{fig:CDForig}
\includegraphics[scale=0.5]{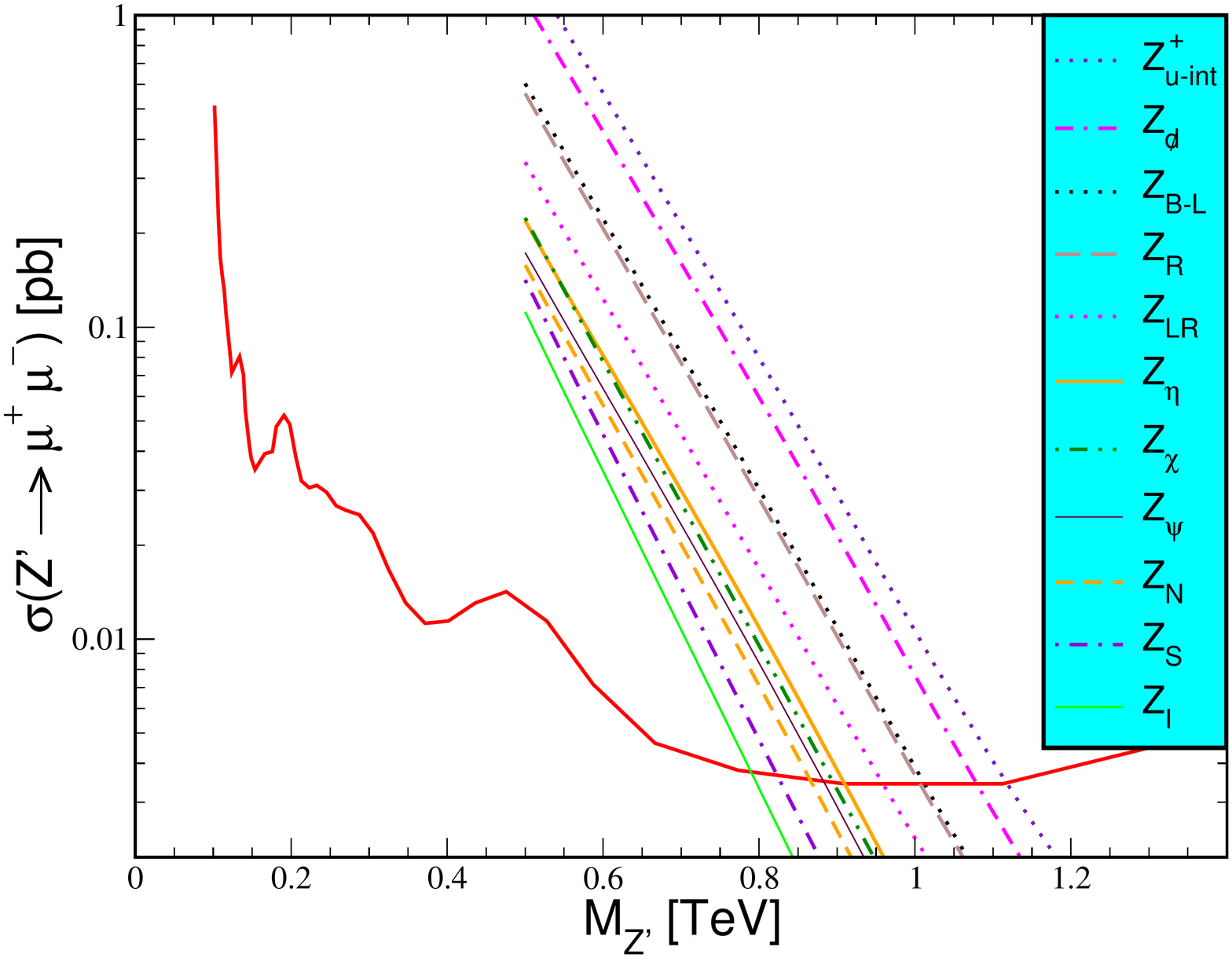}
\caption[]{The solid red curve corresponds to the CDF 95\% C.L. upper limit on the DY production cross-section of a $Z'$ boson as a function of its mass. 
The intersection points with the theoretical cross-sections $\sigma_{Z'}^{\sbx{0.6}{LO-PDF}}$ for various models 
using Eq.~(\ref{eq:lo3}) and CTEQ6L PDFs give the corresponding $M_{Z'}$ lower bounds.
See Table~\ref{tab:limits} for numerical values obtained by this procedure.}}

The theoretical production cross-section of a $Z'$ boson at a hadron collider depends on certain crucial factors, such as the treatment 
of the parton distribution functions (PDFs) of the ingoing quarks and of the radiative corrections to the leading order (LO) process. 
The PDF sets for quarks and gluons are evaluated at various perturbative orders 
for a wide range of factorization scales and momentum fractions by a number of independent groups. 
These sets generally agree with high precision, so that the choice depends on whether a particular group provides PDFs 
at the required perturbative order, the inclusion or neglect of small corrections, the data sets available as of the latest update, {\em etc}. 
For the publication~\cite{Aaltonen:2008ah} we base our analysis on, 
the CDF Collaboration has employed CTEQ6~\cite{Pumplin:2002vw} PDFs.
The PDF sets have since been updated a number of times by the CTEQ group. 
We redo this analysis using the latest sets available and verify the limits using the MSTW set~\cite{Martin:2009iq}.   

As shown in Eq.~(\ref{eq:nlo}) of the appendix, 
for every parton the next-to-leading order (NLO) differential Drell-Yan (DY) cross-section consists of three main parts: 
the PDFs for the incoming hadrons, the parton-level hard cross-section and the QCD higher order terms.
The determination of the PDFs requires experimental input. 
To evaluate them, the parameters of some functional form are fit to the data sets from a number of experiments 
(see, \eg Ref.~\cite{Morfin:1990ck}). 
The central fit, $S_0$, corresponds to the minimum of the $\chi^2$ function.
To allow error estimates the CTEQ and MSTW Collaborations also provide PDF sets, $S_i^{\pm}$,
which are defined as the eigenvectors of the Hessian matrix~\cite{Pumplin:2001ct}.
Thus, the $S_i^{\pm}$ are uncorrelated by construction, providing an efficient method of calculating the induced variations
of the PDF predictions for a chosen practical tolerance value, $T$, defining the region of `acceptable fits' with $\Delta \chi^2\le T^2$. 
The eigenvectors of the Hessian matrix are normalized in such a way that the confidence levels correspond to hyper-spheres.
The uncertainty can then be computed from the simple master formula~\cite{Pumplin:2002vw},
\be
\label{eq:eigen}
\Delta X = \frac{1}{2} \Bigg[ \sum_{i=1}^{N_p} \left[ X(S_i^+)-X(S_i^-) \right]^2\Bigg]^{1/2},
\ee
where $N_p$ is the number of eigenvectors, $X$ is the observable (in our case the $Z'$ cross-section $\sigma_{Z'}$) 
and $X(S_i^\pm)$ are the predictions for $X$ based on the PDF sets $S_i^{\pm}$.

\TABLE[t]{\label{tab:limits}
\begin{tabular}{|l|r|r|r|r||}
\hline
$Z^{\prime}$& this work & CDF & electroweak & projection \\ \hline
$Z_\chi$       &895  &892 &1141 &963\\
$Z_\psi$       &883  &878 &147  &965\\
$Z_\eta$       &910  &904 &427  &984\\
$Z_I$          &789  &789 &1204 &857\\
$Z_N$          &865  &861 &623  &949\\
$Z_S$          &823  &821 &1257 &896\\
$Z_R$          &1006 &    &442  &1071\\
$Z_{B-L}$      &1012 &    &546  &1088\\
$Z_{LR}$       &959  &    &998  &1012\\
$Z_{\not d}$   &1079 &    &472  &1137\\
$Z_{u-int}^{+}$&1117 &    &762  &1182\\
$Z_{SM}$       &1030 &1030&1403 &1076\\\hline
\end{tabular}
\caption{95\% C.L. limits on the masses of some benchmark $Z'$ models.
Given in the first column are the limits which we obtain following the CDF approach as illustrated in Fig.~\ref{fig:CDForig}.
We used here CTEQ6L PDFs in order to be able to directly compare our results with those published by 
the CDF Collaboration~\cite{Aaltonen:2008ah} which are shown in the second column.
The third column contains the limits obtained from the electroweak precision data~\cite{Erler:2009jh}. 
Finally, 95\% C.L. limits projected for an expected integrated luminosity of 8~fb$^{-1}$ are listed in the last column.}}

CDF used CTEQ6 PDFs for their calculation of the DY cross-section for
which CTEQ employed the Particle Data Group average for the strong coupling, 
$\alpha_s(M_Z)= 0.118$~\cite{Nakamura:2010zzi}.
The CTEQ6M package contains the LO sets in addition to the central NLO PDF sets as well as the eigenvector sets for the latter. 
The latest version 6.6M~\cite{Nadolsky:2008zw} has 22 pairs of eigenvector sets for error calculations. 
In the PDF sets produced by the MSTW group, $\alpha_s$ has been treated as a fit parameter. 
This results in LO, NLO, and NNLO $\alpha_s(M_Z)$ values of approximately 0.139, 0.120 and 0.117, 
respectively~\cite{Martin:2009iq,Martin:2009bu}. 
The MSTW sets contain LO, NLO and NNLO PDFs, in each case along with 20 pairs of eigenvector sets.

The QCD corrections to the LO hadronic process of $Z'$ production may considerably alter the magnitude of the cross-section. 
Conventionally, these corrections are taken into account with a `$K$-factor', labelled here as $K^{m/n}$, and is defined as,
\be
\label{kfactor}
K^{m/n} = \frac{d\sbx{1.2}{$\sigma$}^{\sbx{0.7}{$N^mLO$}}}{dM}\left(\frac{d\sbx{1.2}{$\sigma$}^{\sbx{0.7}{$N^nLO$}}}{dM}\right)^{-1},
\ee
where $\sbx{1}{$\sigma$}^{\sbx{0.6}{$N^mLO$}}$ is the differential cross-section to order\footnote{While it may be obvious from the 
definition~(\ref{kfactor}), we recall that the cross-sections need to be evaluated with the corresponding order PDFs.} $m$ in $\alpha_s$.
This factor expresses higher order corrections to $Z'$ production only and not to the complete process. 
If we consider only one quark flavor in Eq.~(\ref{kfactor}) then $K^{m/n}$ is independent of the $Z'$ model and  $M_{Z'}$.
Thus, the proper way to account for higher order QCD corrections is to calculate a different $K^{m/n}$ for every flavor, 
even though it is common practice to choose a universal factor for all flavors and models~\cite{Carena:2004xs}. 
The main results of the present paper are calculated for $\sigma_{Z'}^{\sbx{0.6}{NLO-PDF}}$
defined in Eq.~(\ref{eq:lo3}), with a factor $K^{2/1}$ included as we now explain.

\FIGURE[t]{\label{fig:CDForig2}
\includegraphics[scale=0.5]{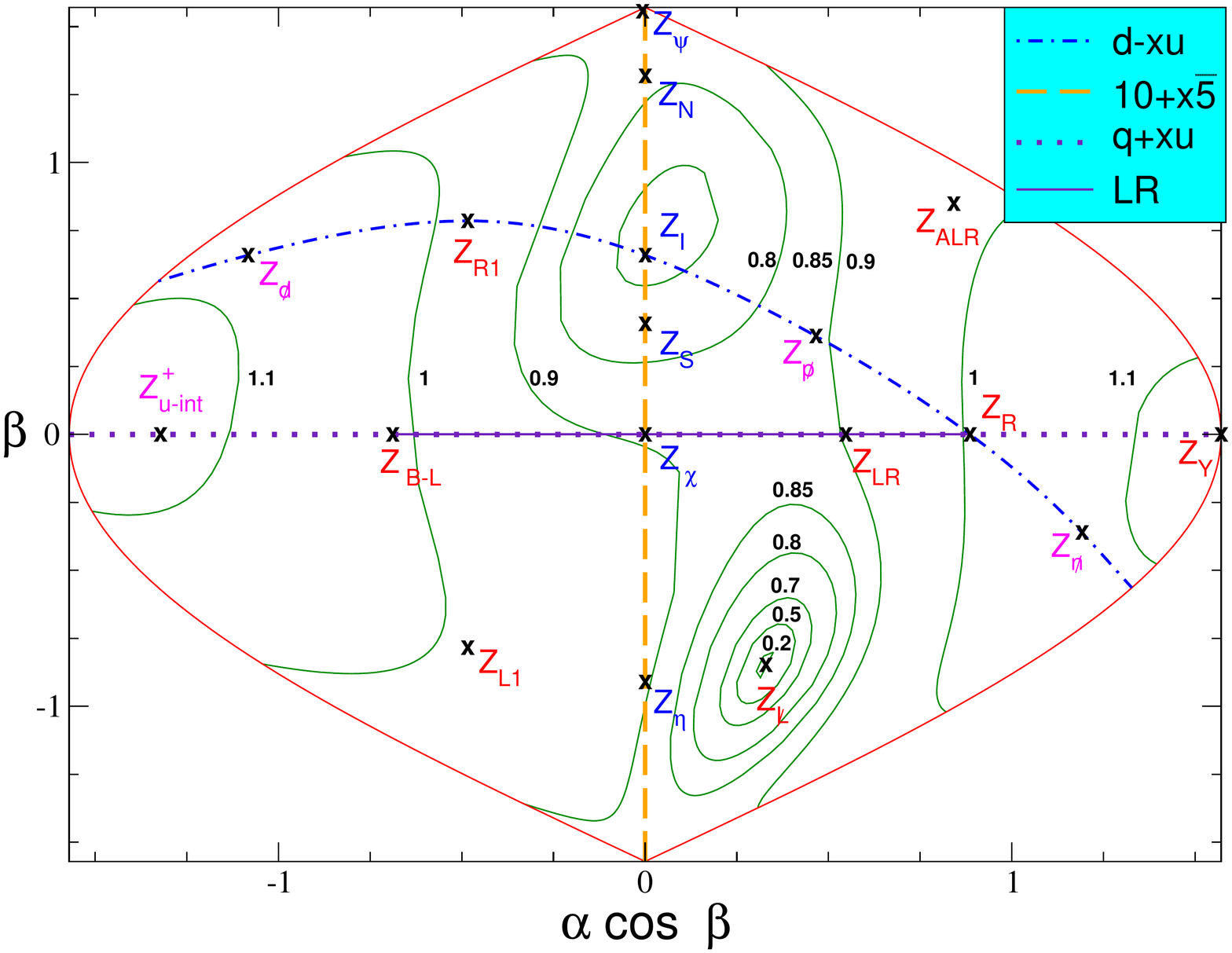}
\caption[]{$\alpha$-$\beta$ Sanson-Flamsteed projection of $E_6$ $Z'$ models.
The contours show the $95\%$~C.L. $M_{Z'}$ limits in TeV 
which are from the intersection of $\sigma_{Z'}^{\sbx{0.6}{LO-PDF}}$ with the CDF 95\%~C.L. upper limit on the cross-section. 
The dotted, dashed, dot-dashed  and indigo colored continuous lines correspond to one-parameter models. 
For this plot we use CTEQ6L PDFs.}}

As intimated earlier, CDF has used a LO expression (with LO PDFs) and a $K^{2/0}$-factor 
(taken from Ref.~\cite{Carena:2004xs}) for their calculation of the DY cross-section via $Z'$ exchange. 
We have compared the $K^{1/0}$ factors from that work with effective $K^{1/0}$ values obtained using 
the expressions given in the appendix~\ref{app:formalism} and found good agreement (within a few~\%). 
Then we also adopted $K^{2/1}$-factors in such a way that our results are effectively NNLO. 
Only for the comparisons in Table~\ref{tab:limits} with the CDF results we use the factor $K^{2/0}$ instead\footnote{For 
invariant masses beyond 1~TeV at the Tevatron, we use constant factors $K^{2/0} = 1.302$ and $K^{2/1} = 1.137$.}.
As for the LHC, NLO results suffice~\cite{Petriello:2008zr} and we take  $K^{2/1} = 1$. 

We take the $95\%$ cross-section upper limit from the data curve in Fig.~3 of Ref.~\cite{Aaltonen:2008ah} and 
find the intersection with our cross-section, $\sigma_{Z'}^{\sbx{0.6}{LO-PDF}}$, as defined in Eq.~(\ref{eq:lo3}) for $Z'$ boson exchange.
Final state radiation effects~\cite{Baur:1997wa} can be ignored in this case since we integrate over almost all invariant
mass range in such a way that these effects are mostly canceled according to the Kinoshita theorem~\cite{Kinoshita:1962ur}. 
The evaluation of the multi-dimensional integrals involved is done 
using the CUHRE and SUAVE programs under the CUBA package~\cite{Hahn:2004fe}. 
We input the numerical values of couplings and charges from the FORTRAN package GAPP~\cite{Erler:1999ug}.
Only $Z'$ decays into SM fermions are assumed and fermion masses neglected. 
A numerical routine extracted from the PEGASUS package~\cite{Vogt:2004ns} is used for the running of $\alpha_s$ from its value 
at $M_Z$ to the factorization scale $M$. 
Fig.~\ref{fig:CDForig} shows the `model-lines' for the $Z'$ bosons including some models not included in the original CDF analysis. 
The slopes of the model-lines and their intersection points with the experimental data line (giving the $M_{Z'}$ limits) 
match with those in the original CDF plot within a few per mille for models included in both analyses.
The 95\% C.L. mass limits for various models are listed in Table~\ref{tab:limits}. 
For comparison, the CDF limits from~\cite{Aaltonen:2008ah} and the EWPD limits from~\cite{Erler:2009jh} are also quoted. 
The last column in the table gives the 95\% C.L. mass limits anticipated at the end of the current Tevatron run, obtained using the Bayesian statistical method explained in the next section.
 
\TABLE[t]{\label{tab:PDFunct1}
\begin{tabular}{|l|r|r|r|}
\hline
$Z^{\prime}$ & CTEQ6M~\cite{Pumplin:2002vw} & CTEQ6.6M~\cite{Nadolsky:2008zw} & MSTW2008~\cite{Martin:2009iq} \\ \hline
$Z_\chi$        & 885  &  886 &  875 \\
$Z_I$             & 779  &  783 &  764 \\
$Z_{\not{d}}$ &1070 &1068 &1063 \\ \hline 
\end{tabular}
\caption{95\% C.L. limits on $M_{Z'}$ obtained using three different NLO PDFs.}}

Fig.~\ref{fig:CDForig2} shows the area preserving sinusoidal (Sanson-Flamsteed) projection of a hemisphere 
parameterizing the $Z'$ bosons in terms of the $E_6$ angles $\alpha$ and $\beta$.  
We note that the $M_{Z'}$ lower bound for $Z_{LR}$ quoted in Table~\ref{tab:limits}
corresponds to the normalization given in Ref.~\cite{Langacker:2008yv}, 
and hence differs from the corresponding value in Fig.~\ref{fig:CDForig2},
where all the models are $E_6$ normalized as in Eq.~(\ref{eq:LR}).

In Table~\ref{tab:PDFunct1} we display the dependence of these limits on the PDFs. 
To also investigate the uncertainties due to them we use the central CTEQ6M PDFs and the corresponding eigenvector PDF sets
as displayed in Table~\ref{tab:errors} for two selected values of $M_{Z'}$.
One sees that the relative uncertainty in $\sigma_{Z^{\prime}}$ is very large for the $Z_I$ 
which is due to $Q'_u = Q'_{\bar{u}} = 0$ in this model 
and points to the fact that for a given $M$ the uncertainties in the $d$ quark PDFs are larger than those of the $u$ quarks. 
Also, the $d$ quark contribution is suppressed by more than an order of magnitude with respect to that of the $u$ quark 
which is reflected by the ratio of the cross-sections for the $Z_{I}$ and the $Z_{\not d}$. 
We recall that we use a common normalization for all models so that the cross-sections can be directly compared. 
Finally we show in the table how the uncertainty in $\sigma_{Z'}$ affects the $M_{Z'}$ limits in these models.

\section{The Bayesian statistical method}
\label{sec:Bayes}
The CDF Collaboration collected an integrated 2.3~fb$^{-1}$ of data~\cite{Aaltonen:2008ah} in the $\mu^+\mu^-$ channel,
binned in inverse invariant di-muon mass, $m_{\mu\mu}^{-1}$. 
The CDF analysis then looks for an enhancement in di-muon production above the SM background for
particular $E_6$ models, and so their lower limits on $M_{Z'}$ correspond to upper limits on the cross-section\footnote{It utilizes 
signal templates that have been generated with a fixed and relatively narrow $\Gamma_{Z'} = 2.8\% \times M_{Z'}$ 
(motivated by $\Gamma_Z$).}.
But $Z'$ bosons interfere with the SM neutral gauge bosons, and destructive interference would result 
in a reduction of the SM cross-section.
For this reason in addition to the general motivation given in Section~\ref{sec:intro} 
--- clear-cut combination of $Z'$ constraints from quite distinct sources ---
we adopt a statistical framework wherein it is 
straightforward to address interference effects
and to vary the coupling strength (see also Ref.~\cite{Sert:2010ma}) up to the strong coupling regime (and broad resonances). 

\TABLE[t]{\label{tab:errors}
\begin{tabular}{|c||c|r||c|r||r||}
\hline
& \multicolumn{2}{c||}{$M_{Z'} = 0.8$~TeV} & \multicolumn{2}{c||}{$M_{Z'} = 1.1$~TeV} & \\
\hline
$Z'$ & $\sigma_{Z'}$~[fb] & $\Delta\sigma$ & $\sigma_{Z'}$~[fb]&$ \Delta\sigma$ & $M_{Z'}$~[GeV] \\ 
\hline\hline
$Z_{\chi}$    & $8.6 \pm0.7$   &  8\% & $  0.41\pm 0.04     $&9\% &$885 \pm 9$   \\
$Z_I$         & $3.0 \pm0.7$   & 24\% & $0.14\pm0.03$&20\%  &$779_{-25}^{+20}$ \\
$Z_{\not d}$  & $54 \pm4$  &  8\% & $2.6 \pm0.3 $&11\%  &$1070_{-12}^{+11}$ \\\hline
\end{tabular}
\caption[]{Di-muon cross-section for $Z'$ exchange obtained using NLO CTEQ6M PDFs sets for two representative values 
of $M_{Z'}$, along with the uncertainty due to the eigenvector PDF sets. 
The percentages of the uncertainties are given in the third and fifth columns. 
In the last column we give the 95\% C.L. limits on $M_{Z'}$ and the positive and negative uncertainties in these. 
We obtain the uncertainties from the intersection of 
$\sigma^{\sbx{0.6}{NLO-PDF}}_{Z'} \pm \Delta\sigma^{\sbx{0.6}{NLO-PDF}} _{Z'}$ 
for $Z'$ exchange and using Eq.~(\ref{eq:eigen}) with the CDF 95\% C.L. upper limit on the cross-section.}}

The basic idea is to apply the Bayesian analysis of the SM Higgs mass, $M_H$, of Ref.~\cite{Erler:2010wa} to $Z'$ physics. 
In this case the collider constraints from LEP~2~\cite{Barate:2003sz} and the Tevatron~\cite{Aaltonen:2010yv} 
were included using the published log-likelihood ratios,
\be
   {\rm LLR}_i \equiv -2 \ln {\cal L}_i \equiv -2 \ln {p({\rm data}|s+b)\over p({\rm data}|b)}.
\ee
These are given in terms of the probabilities (likelihoods) to obtain the data, conditional on the signal plus background hypothesis, 
$p({\rm data}|s+b)$, and background only hypothesis, $p({\rm data}|b)$, and may be compounded of many experiments, channels,
energies, {\em etc.} The ${\cal L}_i$ depend on the parameter(s), $\mu$, of interest, ($\mu = M_H$ in Ref.~\cite{Erler:2010wa}), 
through the signal hypothesis. Information on $\mu$ is obtained by Bayes's theorem,
\be
\label{eq:likelihood}
   p(\mu|{\rm data}) = {p({\rm data}|\mu) p(\mu) \over p({\rm data})},
\ee   
where $p(\mu)$ is the prior probability density function (pdf) entering Eq.~(\ref{eq:posterior}),
and is a summary of our knowledge, if any, prior to the experiment or analysis. 
In the absence of prior information, or if the prior information is explicitly taken into account by extra factors $p(\mu|n_i)$, 
then $p(\mu)$ is called {\em non-informative}, and is most conservatively taken as $p(\mu) = 1$ or $p(\mu) = \mu^{-1}$, 
whenever $\mu$ may be an arbitrary real number or positive real number, respectively.
Notice, that $p({\rm data})$ drops out from likelihood ratios. 
This is crucial: if various data points show poor compatibility, or if an excess (or deficit) is observed, 
this will have an impact only if some value of $\mu$ describes the data better than some other. 

\FIGURE[t]{\label{fig:PDFs}
\includegraphics[scale=0.4]{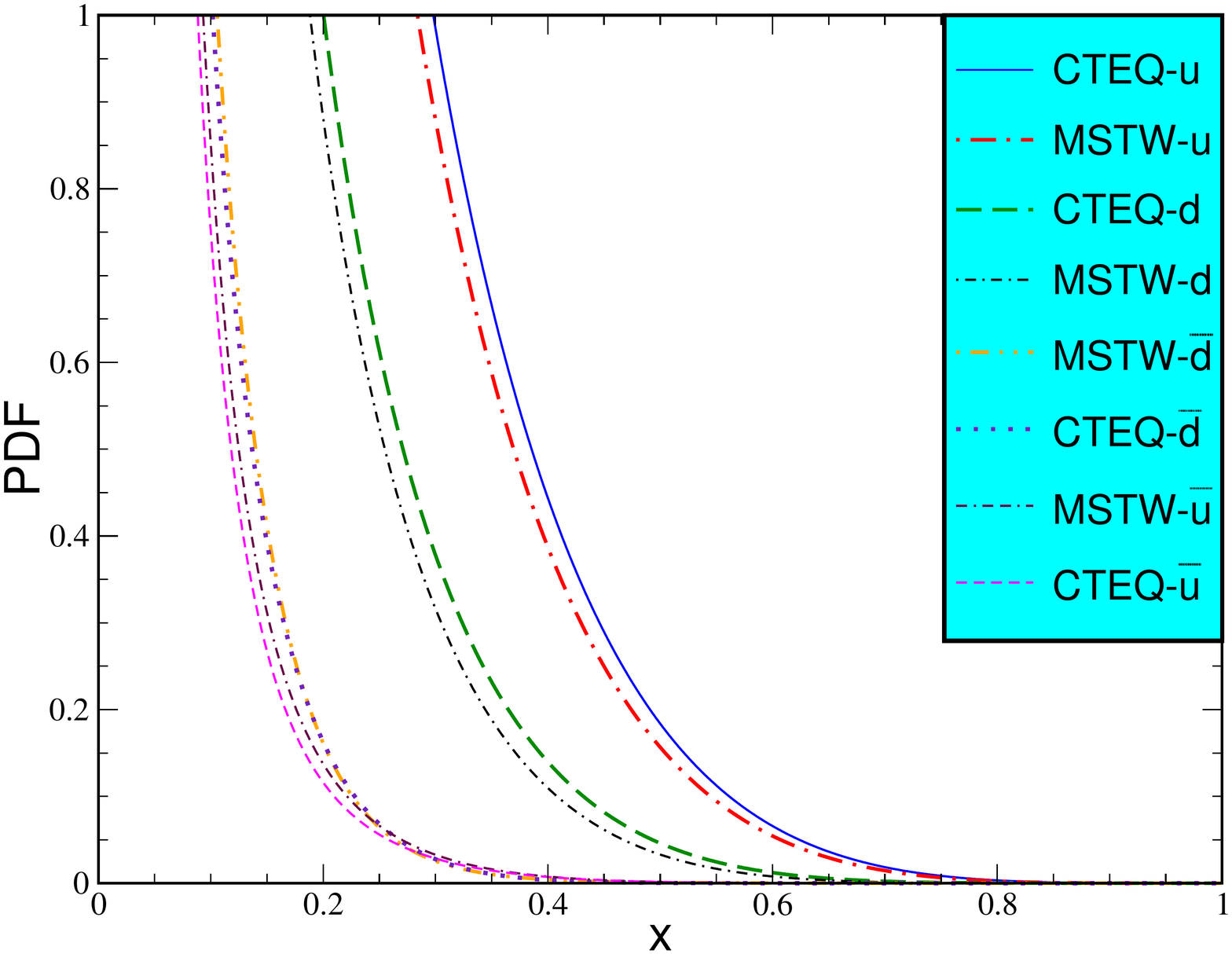}
\includegraphics[scale=0.4]{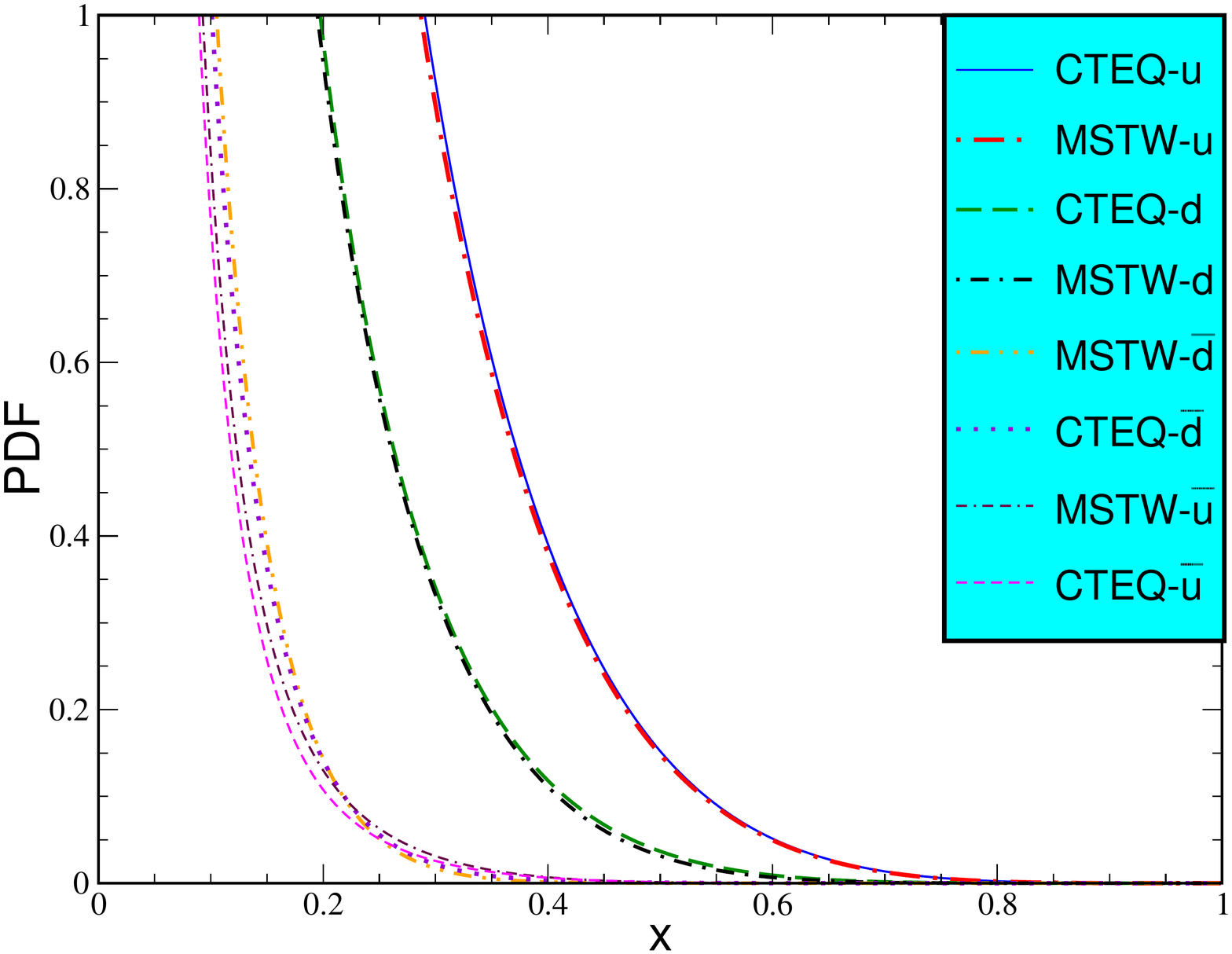}
\caption{The $x$ dependence of the (proton) PDFs at $Q^2 = M^2 =(1000$~GeV)$^2$. 
The upper (lower) plot shows the LO (NLO) PDFs of $u$, $d$, $\bar{u}$, and $\bar{d}$ quarks 
provided by the CTEQ and MSTW groups.}}

We use here a similar philosophy as in Ref.~\cite{Erler:2010wa}, 
but here we first have to construct the corresponding LLR ourselves.
Our input data are the number of events, $n_i$, in bin $i$ (see Fig.~1 in Ref.~\cite{Aaltonen:2008ah}).
The parameter set $\mu$ may include all of the parameters introduced in Sections~\ref{sec:intro} and \ref{sec:models},
namely $M_{Z'}$, $g'$, $\theta_{ZZ'}$, $\alpha$, and $\beta$, but in this paper we allow only $M_{Z'}$ and $g'$ for
some specific models (\ie fixed values of $\alpha$ and $\beta$) and set $\theta_{ZZ'} = 0$.
Complementary data sets, such as other channels, LHC and D\O\ results, EWPD data~\cite{Erler:2009jh,delAguila:2010mx}, 
and LEP~2 constraints~\cite{Alcaraz:2006mx}, will be necessary to disentangle these parameters in an integrated analysis.
The SM point corresponds to $M_{Z'}^{-1} = 0$ or $g' = 0$. The events, $n_i$, in each bin follow Poisson statistics,
\be
\label{eq:poisson}
   P(n_i|\nu_i) = {\nu_i^{n_i} e^{-\nu_i} \over n_i!},
\ee
where $\nu_i$ is the predicted number of events in bin $i$ given specific values for $M_{Z'}$ and $g'$. 

It is important to note here that the above likelihood is the same as the one employed 
by the CDF collaboration in their analysis~\cite{Aaltonen:2007ps}. However, they determine $\nu_i'$ by summing the bin-counts 
from the SM and the bin-counts from the signal template as mentioned earlier, effectively summing the 
cross-sections of the SM process and the $Z'$-mediated process without any interference. Therefore, the CDF approach 
 essentially differs from ours not in the choice of the likelihood, but in that the interference 
effects were neglected in their analysis in order to keep it simple and model-independent~\cite{AK:2011}, 
deliberately making it blind to wider resonances through the use of templates based on narrow signal width. 
We, conversely, treat coupling strength as a free parameter and avoid signal templates, which makes the inclusion of 
interference effects rather natural in our framework.

In practice, we compute a grid\footnote{Alternatively calculating $\chi^2$ {\em directly\/} without the grid gives mass limits 
which differ by at most 3~GeV for our benchmark models.
However, this dramatically increases CPU time if a multi-variate minimization
is performed \ie without fixing $g'$ and other model parameters.
It is therefore expedient to avoid the mostly redundant PDF integrations.}
of values for the $\nu_i$ discretizing  $M_{Z^{\prime}}$ and $g'$ and interpolate in every one of the 
first 35 bins corresponding to the invariant mass range searched by CDF.
We have thus effectively reduced the analysis to a least-$\chi^2$ fit where any observed event count adds a piece,
\be
\label{eq:chimin}
   \Delta\chi_i^2 = {\rm LLR}_i = 2 \left( \nu_i' - \nu_i + n_i\ln \frac{\nu_i}{\nu_i'} \right),
\ee
to the overall $\chi^2$ function ($\nu_i$ and $\nu_i'$ refer here to the SM and SM plus new physics expectations, respectively).

The detector resolution, $\Delta = 0.17$~TeV$^{-1}$, is approximately constant (in the variable $m_{\mu\mu}^{-1}$),
and must be taken into account since it is of the order of the bin size of (3.5~TeV)$^{-1}$.
We define $M_{\mu\mu}$ as the theoretical invariant mass of the muon pairs,
as opposed to the nominally measured $m_{\mu\mu}$, and introduce the convolution,
\be
\label{eq:bgevents}
\nu_i = \epsilon \left[ \mathcal{L} \int_{\rm bin} d m_{\mu\mu}^{-1}\, A(m_{\mu\mu}) \int\limits_{0}^{\infty} d M_{\mu\mu}^{-1}\, 
p(m_{\mu\mu}^{-1}|M_{\mu\mu}^{-1}) \, K^{2/1}K_{\gamma}{d\sigma^{\sbx{0.6}{NLO}}\over dM_{\mu\mu}^{-1}}+\nu_{n\scalebox{0.5}{DY}} 
\right],
\ee
where $\mathcal{L} = 2.3$~fb$^{-1}$ is the integrated luminosity, 
$\epsilon$ is the detector efficiency, which we take as a constant 0.982 for all bins, 
$\sigma^{\sbx{0.6}{NLO}}$ is given by Eq.~(\ref{eq:nlo}), 
and $\nu_{n\scalebox{0.5}{DY}}$ refers to the non-DY background which is extracted directly from Fig.~1 of Ref.~\cite{Aaltonen:2008ah}. 
$A(m_{\mu\mu})$ is the total acceptance of the CDF detector, 
which increases from about $0.13$ at the $Z$-pole to about $0.4$ at 1~TeV and then falls off rapidly~\cite{Aaltonen:2008ah}.
For our current analysis, the acceptance values have been gleaned from~\cite{Acceptance}.

\FIGURE[t]{\label{fig:interference}
\includegraphics[scale=0.5]{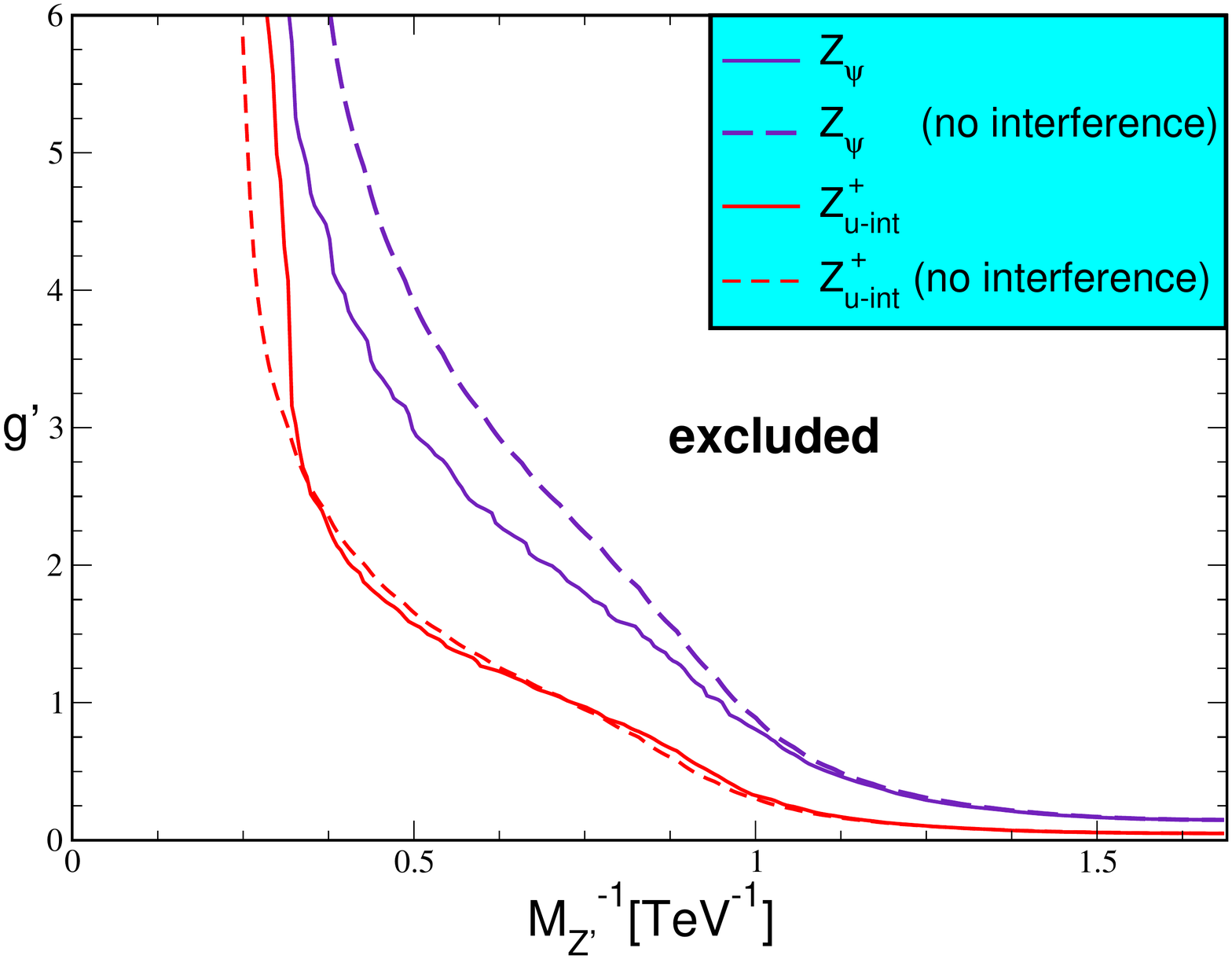}
\caption{The 95\% C.L. exclusion contours in the $g'\; vs.\; M_{Z'}^{-1}$ plane for the $Z_{u-int}^+$ and $Z_\psi$ bosons.
These are compared to the contours that are obtained when the interference effects are ignored.}}

Along with the QCD corrections, the $\mathcal{O}(\alpha)$ QED radiative corrections~\cite{Baur:1997wa} 
also have a sizable effect on the DY cross-section, and strongly affect the shape of the di-lepton invariant mass distribution. 
While initial state radiation is negligible for di-muon masses between 50 and 100~GeV, 
final state QED corrections are in fact larger than the $\mathcal{O}(\alpha_s)$ QCD corrections and so have to be taken into account.
The QED corrections are shown in Fig.~6 of Ref.~\cite{Baur:1997wa} with a rapid variation visible in the range 
$40~{\rm GeV} < M < 110$~GeV. 
Just below the $Z$ peak, these corrections enhance the cross-section by up to a factor of 1.9,  
so the cross-sections in the neighboring bins to the $Z$ peak differ considerably from the values expected without these corrections. 
When the full di-muon mass range is integrated over, the large negative and positive corrections tend to cancel and do 
not have a big impact on the total cross-section~\cite{Kinoshita:1962ur}.
For $M >130$~GeV, they uniformly reduce the differential cross-section by 7\%, and for our 
calculations\footnote{The figure~\cite{Baur:1997wa} has been generated for $\sqrt{s}=1.8$ TeV,  while our process is being computed 
for the $\sqrt{s}=1.96$ TeV of the current Tevatron run, but we expect this to have negligible effect on our final results.} 
we have extrapolated the data points from the mentioned figure and used these as multiplicative factors, 
which we refer to as $K_{\gamma}(M_{\mu\mu})$.
In principle, such effects should also appear near the $Z'$-pole, but considering that the bin around the expected $Z'$ mass is fairly wide,
any large effect around the peak will be washed out at least for weak and intermediate coupling strength.

Returning now to Eq.~(\ref{eq:bgevents}), the quantity $p(m_{\mu\mu}^{-1}|M_{\mu\mu}^{-1})$ is our smearing function,
\be
   p(m_{\mu\mu}^{-1}|M_{\mu\mu}^{-1}) \equiv m_{\mu\mu}\, {b^a  e^{-b} \over \Gamma(a)},
\ee
with $a^{-1} = M_{\mu\mu}^2\, \Delta^2$ and $b^{-1} = m_{\mu\mu}\, M_{\mu\mu}\, \Delta^2$ and
is constructed as a Beta distribution with mean $M_{\mu\mu}$ and variance $\Delta^2$.
Note that it approaches a Gaussian form for $a \gg 1$ (which is the case except for the first few bins), 
but is more adequate than a Gaussian since $M_{\mu\mu}$ takes non-negative values only. 
We note in passing that we neglect here systematic and theoretical uncertainties,
justifying this with the very small event numbers in the most relevant bins so that statistics dominates.

\FIGURE[t]{\label{fig:chimodel}
\includegraphics[scale=0.5]{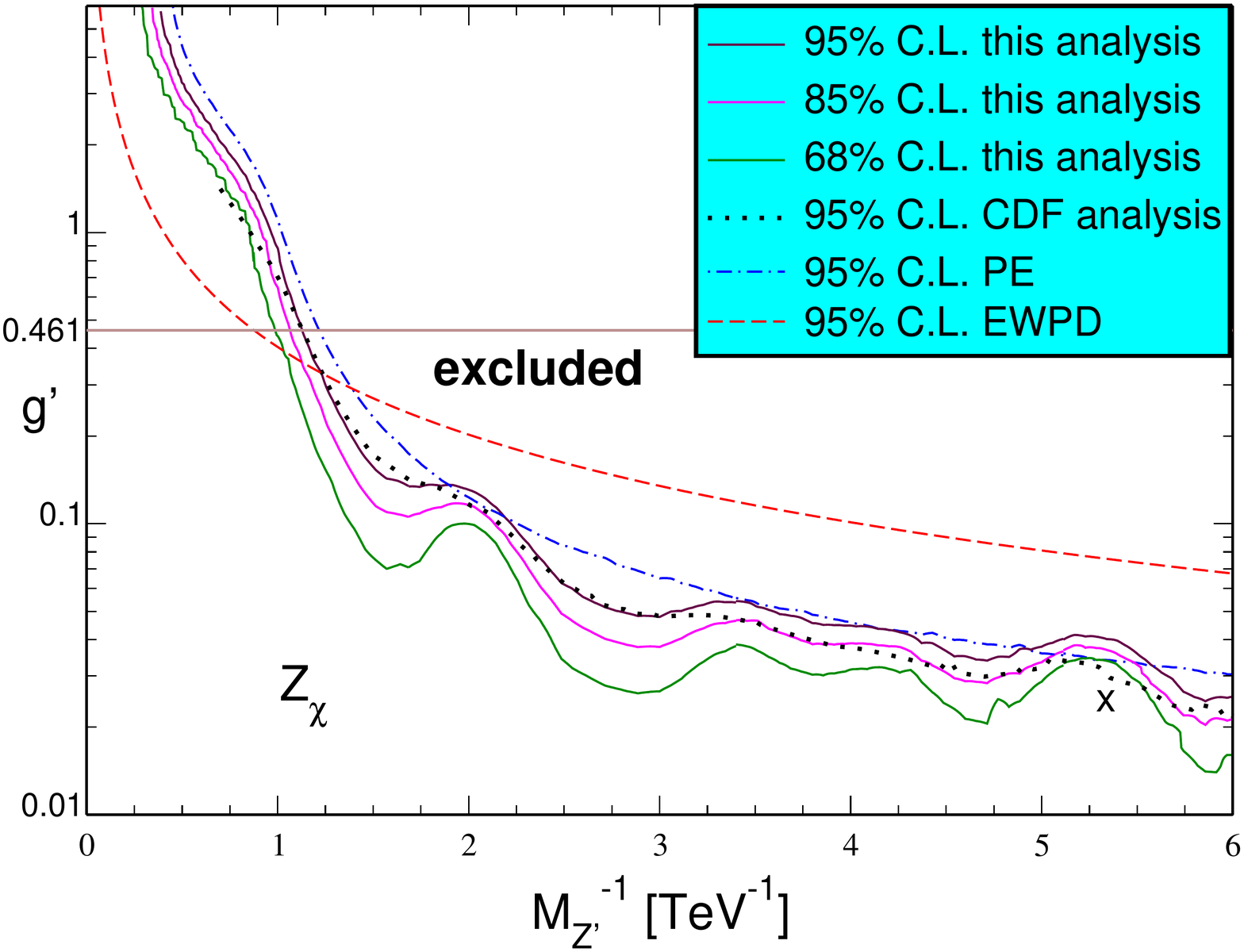}
\caption{Contours in the $g'\; vs.\; M_{Z'}^{-1}$ plane for the $Z_\chi$ model. 
The solid lines correspond --- from top to bottom --- to the 95\%, 85\%, and 68\% C.L. contours using the Bayesian analysis,
and are to be compared with the dotted line using the CDF (frequentist) approach.
The dot-dashed line shows what one would expect from the Bayesian method based on pseudo-experiments (PE),
while the EWPD yield the dashed line.
The horizontal line indicates the value $g' = 0.461$ which is motivated by gauge coupling unification and often used as reference. 
The cross indicates the best fit value.}}

\FIGURE[t]{\label{fig:results}	
\includegraphics[scale=0.5]{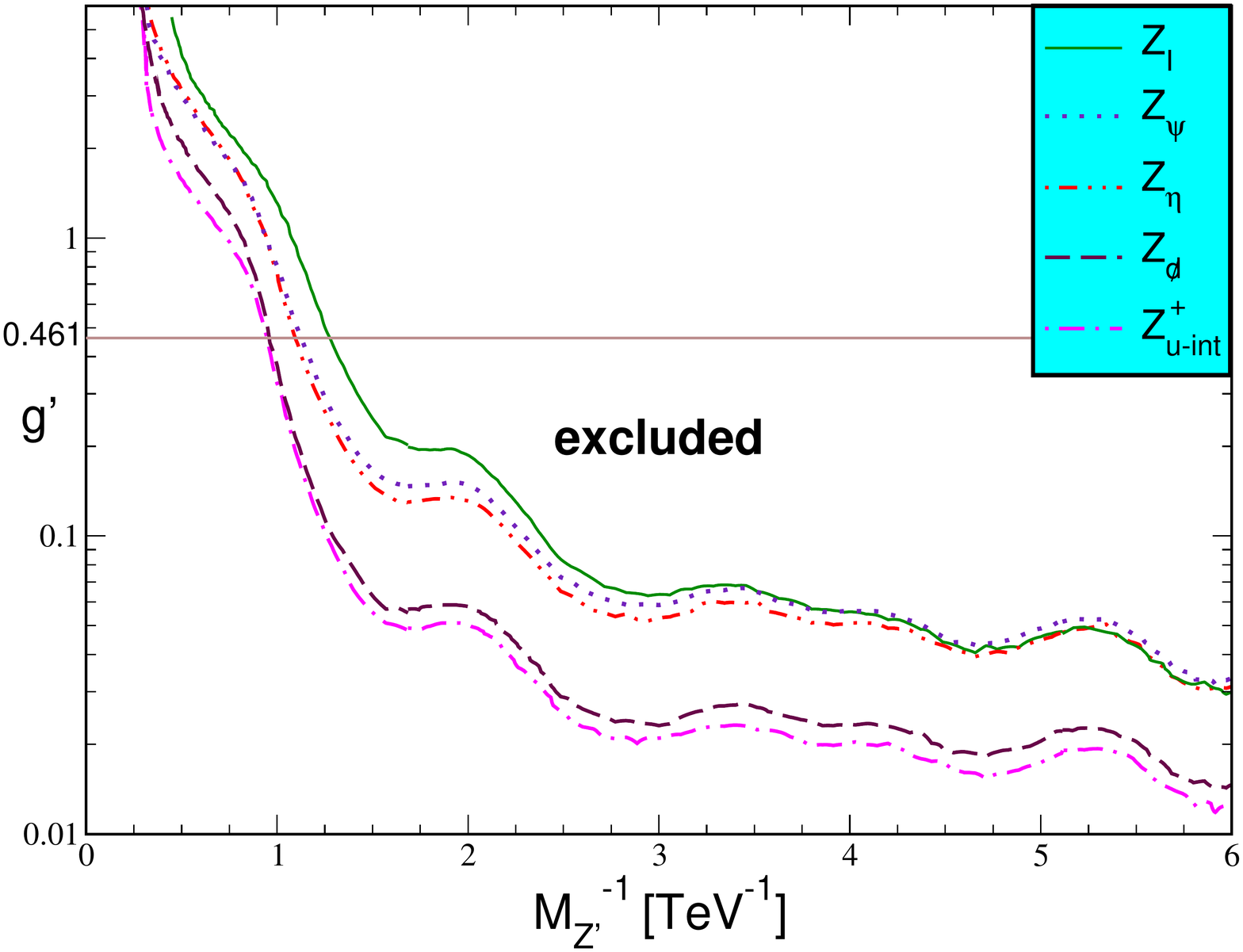}
\caption{95\% C.L. contours  in the $g'\; vs.\; M_{Z'}^{-1}$ plane for five representative $E_6$ inspired models 
using our Bayesian method.}}

Eq.~(\ref{eq:chimin}) also makes it explicit how our approach allows the new physics to enter with either sign, 
as is always the case for interfering $Z'$ bosons.
At the level of the differential cross-section, 
the interference terms change sign when $M_{\mu\mu}$ crosses the $Z$ or $Z'$ poles.
Thus, there are fairly large cancellations at work when the whole range of $m_{\mu\mu}$ is integrated over, 
and when the objective is the usual hunting for a narrow bump where neglect of interference effects is justified \cite{Hays:2009dh}. 
Since here we put more emphasis on the event distribution over larger numbers of bins, the interference issue becomes more interesting.
To have a closer look as to how significant the interference effects are numerically, 
we now discuss two cases where they are enhanced, $Z_{u-int}^+$ and $Z_{u-int}^-$. 
They are defined to have, respectively, maximum constructive and destructive interference with the SM amplitudes for up quarks 
in the limit $M_{\mu\mu} \to \infty$, \ie we extremize the expression (in a slightly more compact notation),
\be
\label{inter2}
   e^2 Q_u Q_\mu (Q'_{Lu} + Q'_{Ru}) (Q'_{L\mu} + Q'_{R\mu}) +
   g_Z^2 (\epsilon_{Lu} Q'_{Lu} + \epsilon_{Ru} Q'_{Ru}) (\epsilon_{L\mu} Q'_{L\mu} + \epsilon_{R\mu} Q'_{R\mu}).
\ee
We chose $M_{\mu\mu}^{-1} = 0$ as our reference value because then any dependence on $M_{Z'}$ and $\Gamma_{Z'}$ drops out.
Moreover, at large $M_{\mu\mu}$ the down quark contribution to the PDFs is strongly suppressed, providing a further simplification.
For the case of the $Z_{u-int}^+$ it now turns out that neglecting the second ($ZZ'$ interference) term 
shifts the corresponding values for our model parameters $\alpha$ and $\beta$ only at the $10^{-2}$ level, 
so that we can neglect this term, as well, and we find,
\be\label{intersol}
   \alpha = {1\over 2} \arctan \frac{2\sqrt{6}}{9} - {\pi\over 2} \approx -76^\circ, \hspace{24pt} 
   \beta= 0,
\ee
which is relatively close to the $Z_{B-L}$ case.  
The facts that the $Z_{B-L}$ boson couples only vector-like and that the vector-coupling for the muons is suppressed by a factor 
$1 - 4 \sin^2\theta_W(M_{Z'}) \approx 0.04$, may give a rationale for why in this case the $ZZ'$ interference term is small.
Similarly, the $Z_{u-int}^-$ is numerically close to the $Z_\psi$ boson which has only axial-vector couplings to the SM fermions. 
In this case we can neglect the first ($\gamma Z'$ interference) term in Eq.~(\ref{inter2}) and simply define $Z_{u-int}^- \equiv Z_\psi$.
As for the integrated cross-sections, the constructive interference for $Z_{u-int}^+$ is about an order of magnitude larger 
than the destructive interference in $Z_{u-int}^-$, 
and in the latter case we find that the sign of the interference effect in the total cross-section is reversed compared to the amplitude level
in the $M_{\mu\mu} \to \infty$ limit.

We illustrate the interference effects for the $Z_{u-int}^+$ and $Z_\psi$ bosons in Fig.~\ref{fig:interference}.
As can be seen, they become significant for $g'$ values of order unity.
In fact, for large $M_{Z'} \gtrsim {\cal O}(1~{\rm TeV})$ most of the expected signal events come from the $\gamma Z'$ interference,
since the pure $Z'$ exchange contribution is more strongly mass suppressed. 
Another way to quantify the interference effects is to look at the behavior of the best fit location.
{\em E.g.\/}, for the $Z_{\not{d}}$ model (not included in the plot) we found the global best fit at $M_{Z'} = 2.05$~TeV and $g'= 1.14$,
with the value of $\Delta\chi^2 = -1.43$ relative to the SM.
On the other hand, if we turn off the interference effects, the global minimum strongly shifts to $M_{Z'} = 0.189$~TeV and $g'= 0.011$
with $\Delta\chi^2 = -0.97$.

We also stress that the interferences are important if one wants to discriminate between models\footnote{The
importance of interference effects in forward-backward asymmetries was emphasized in Ref.~\cite{Rosner:1995ft}.}. 
{\em E.g.\/}, the $Z_{u-int}^{+}$ and $Z_{\psi}$ models have their global minimum at low mass and weak coupling 
similar to the values above regardless of the interference. 
But the $\chi^2$-minimum becomes deeper in the presence of interference effects,
even though we show in Fig.~\ref{fig:interference} that the mass limits are unaffected at small coupling. 
Moreover, without interference $\chi^2$ is virtually degenerate at the minimum for the three mentioned models,
but this is lifted by the interference effects. 

\TABLE[t]{\label{tab:grid-limits}
\begin{tabular}{|l|r|r|r|}
\hline
$Z'$ & two-parameter & one-parameter &frequentist \\ 
\hline
$Z_\chi$       & 886 & 914 & 885  \\ 
$Z_\psi$       & 888 & 916 & 871  \\ 
$Z_\eta$       & 913 & 939 & 898  \\ 
$Z_I$            & 784 & 809 & 779  \\ 
$Z_\not{d}$   & 1043 &1064 &1070 \\
$Z_{u-int}^+$ & 1056 &1084 &1110  \\
\hline  
\end{tabular}
\caption{95\% C.L. limits on $M_{Z'}$ [in GeV] for six representative $Z'$ bosons.
Shown are the two-parameter (with $g'$ free) and one-parameter (with $g' = 0.461$ fixed) limits. 
In the last column we show the one-parameter limits obtained using the frequentist method as discussed 
in Section~\ref{sec:formalism}.}}

The contours in the $g'\; vs.\; M_{Z'}^{-1}$ plane are given in Fig.~\ref{fig:chimodel} for the $Z_{\chi}$. 
For comparison with our approach, we extended the CDF limit corresponding to $g' = 0.461$ to other values of the coupling.
Crucially, the CDF line breaks down at $g'\sim 1.5$ owing, again, to the fact that their templates
assume a narrow $Z'$ boson. 
As can be seen, our method reveals a strong variation of the $Z'$ mass limit with $g'$, 
while the frequentist method used by CDF shows a weaker and mostly monotonic dependence.
The PE line is obtained by assuming an event count $n_i = \nu_i$ (\ie the SM expectation even though $\nu_i$ is not integer valued) 
in Eq.~(\ref{eq:chimin}), in place of the actually observed number of events for all the bins. 
This yields a smooth and monotonic contour, demonstrating that the strong $M_{Z'}$ dependence of the $g'$ limits is real.
We also show the contour $g' \propto M_{Z'}$ from the EWPD.  
As exemplified here, the EWPD give stronger (weaker) constraints for larger (smaller) values of $g'$ when compared to the CDF data.

At a crude level, the exclusion curves obtained by us using either the CDF method or our own are quite similar, 
and one notices that the downward fluctuations in the cross section data~\cite{Aaltonen:2008ah}
at around 3 and 6~TeV$^{-1}$ are reflected in both cases.  
At a finer level, we tend to see slightly higher $Z'$ mass limits using our Bayesian approach for small $g'$,
compared to those obtained using the CDF approach, while for stronger coupling, $g' \gtrsim 0.5$, the limits are lower for most models.
Our method also maps out fluctuations more faithfully.
These small differences arise from a combination of effects, such as the interference term, the statistical interpretation,
and to a smaller extent our neglect of systematics.
The best fit is also indicated in the figure.  
It is amusing that it occurs for $M_{Z'}$ values close to but somewhat smaller than the invariant masses where 
the CDF Collaboration~\cite{Aaltonen:2008vx} sees a significant deficit (at $m_{ee}\approx 200$~GeV) 
followed by a significant excess (at $m_{ee}\approx 240$~GeV) in di-electron DY production.

As mentioned earlier, it is not possible to extend the CDF limit to values of $g'$ much greater than unity. 
Thus, our approach provides a convenient way to find mass limits for a strongly coupled $Z'$ boson. 
The $M_{Z'}$ limits for a particular $g'$ value can be read off from Figures~\ref{fig:interference}, \ref{fig:chimodel}~and~\ref{fig:results}.
Note, however, that these are two-parameter limits since there are two fit parameters in the minimization of the $\chi^2$ function. 
The 95\% C.L. in this case corresponds to $\Delta\chi^2 = 5.99$, 
while a one-parameter limit with $g'$ fixed corresponds to $\Delta\chi^2 = 3.84$. 
To illustrate this we compare in Tab.~\ref{tab:grid-limits} the values of the two-parameter and one-parameter limits from 
the Bayesian method corresponding to $g' = 0.461$, 
and also show the limits of the (one-parameter) frequentist method for various models.  
As can be seen, the variation of the limits due to different statistical {\em interpretations\/} is larger than any of the theoretical 
uncertainties reviewed in Section~\ref{sec:formalism}.

\TABLE[t]{\label{tab:LHClimits}
\begin{tabular}{|r|rr|rr|rr|rr|}
\hline
 & \multicolumn{2}{c|}{2~TeV} & \multicolumn{2}{c|}{7~TeV} & \multicolumn{2}{c|}{14~TeV} & \multicolumn{2}{c|}{28~TeV} \\ \hline
$\int\mathcal{L}$~[fb$^{-1}$] & $pp$ & $p\bar{p}$ & $pp$ & $p\bar{p}$ & $pp$ & $p\bar{p}$ & $pp$ & $p\bar{p}$ \\ \hline
\hline
3    &0.65  &0.88  &1.60 &2.11   &2.46 &3.26 &3.76 &4.52 \\
10   &0.74  &0.98  &1.86 &2.50   &3.07 &4.05 &4.58 &5.87 \\
30   &0.82  &1.08  &2.14 &2.85   &3.56 &4.52 &5.36 &7.15 \\
100  &0.90  &1.20  &2.43 &3.22   &4.10 &5.29 &6.23 &8.59 \\
300  &0.97  &1.25  &2.69 &3.55   &4.60 &5.97 &7.85 &9.89 \\
1000 &1.05  &1.32  &2.98 &3.89   &5.14 &6.70  &8.98 &11.2 \\
3000 &1.12  &1.38  &3.24 &4.18   &5.63 &7.33 &10.0 &12.5 \\\hline
\end{tabular}
\caption[]{Projected 95\% C.L. exclusion limits [in TeV] on $M_{Z'}$ for the $Z_\chi$ model using our Bayesian method. These limits are obtained by assuming 
the number of observed events $n_{i}$ equal to the SM expectation $\nu_i$ in Eq.~(\ref{eq:chimin}). 
We consider typical CM energies (shown in the top line) 
and a range of integrated luminosities $\int\mathcal{L}$ for $pp$ and $p\bar{p}$ colliders. See the text for details.}}

Finally, we exploit the predictive nature of the Bayesian formalism to give a general idea of how various collider options compare.
For this we project $95\%$ C.L. lower  limits on the mass of the $Z_\chi$ boson for various reference CM energies and luminosities 
(actual and hypothetical) for $pp$ as well as $p\bar{p}$ collisions.
We employed the CDF acceptance (asymptotically for large dimuon masses, we used the constant value of 0.316)
and ignored FSR as well as finite resolution effects in the dimuon invariant mass.
For the resulting limits listed in Table~\ref{tab:LHClimits} we fixed the number of bins to ten, 
and varied the bin size until the mass limit reaches a maximum. 
This optimal bin size turned out to be a constant times the inverse of the corresponding limit.
The cases of 3~fb$^{-1}$ and 10~fb$^{-1}$ correspond roughly to the currently analyzed and final Tevatron data sets,
while 30~fb$^{-1}$ refers to what is often called the low-luminosity LHC. 
Likewise, 300~fb$^{-1}$ and 3000~fb$^{-1}$ correspond to the high-luminosity and luminosity-upgraded LHC, respectively.
The cases of 100~fb$^{-1}$ and 1000~fb$^{-1}$ are included so that they can be compared with the existing projections 
in the literature~\cite{Leike:1997cw,Godfrey:2002tna,Accomando:2010fz}.
The limits in Table~\ref{tab:LHClimits} are much higher for $p\bar{p}$ colliders
where the $\bar{q}$ mostly emerges as a valence quark from the anti-proton with a PDF given by $f^{\bar{p}}_{\bar{u}} = f^p_u$,
while in the case of $pp$ collisions the $\bar{q}$ is always a sea quark and thus has lower PDF values (as illustrated in Fig.~\ref{fig:PDFs}).
Table~\ref{tab:LHClimits} shows that for $Z'$ (DY) physics $p\bar{p}$ colliders have a relative advantage 
as significantly less integrated luminosity is needed to match the $pp$ case,
although larger CM energies tend to mitigate this effect.

\section{Conclusions and outlook}
\label{sec:conclusions}
In this article we have completed a step towards an integrated analysis of the physics parameters associated with extra $Z'$ bosons. 
These parameters,  $M_{Z'}$, $g'$, $\theta_{ZZ'}$, $\alpha$, $\beta$, $etc.$, need to be disentangled and this can be achieved by exploiting 
the complementary nature of electroweak precision data, lepton and hadron colliders, 
as well as the discriminatory power of simultaneously analyzing a variety of processes within each of these data classes. 
The ($\alpha$, $\beta$) parameterization~\cite{Erler:1999nx} discussed in Section~\ref{sec:models} allows us to analyze 
fairly different kinds of models, such as $E_6$ derived models with and without kinetic mixing, 
$Z'$ bosons motivated by chiral models of weak scale supersymmetry~\cite{Erler:2000wu},
models based on left-right symmetry, and other one-parameter models, all on the same footing. 

To move beyond a collection of lower mass limits on the $Z'$ bosons and towards an integrated analysis 
it is crucial to have a common framework.
We proposed such a framework in Section~\ref{sec:Bayes} after having laid out some technical groundwork in 
Section~\ref{sec:formalism}, including a detailed formulation of the DY production of $\mu^+\mu^-$ via neutral gauge bosons 
at hadron colliders such as the Tevatron and the LHC, and a discussion of the associated uncertainties. 
We have shown that the limits obtained with our approach are numerically close to those in the more traditional frequentist approach.
This should not come as a surprise.
Indeed, while the need to introduce a prior probability density is often held against Bayesian data analysis
(a point which has been over-emphasized by many authors), the posterior distributions are often (and certainly here) strongly data driven.
Our resort to Bayes originates more out of the general philosophical mindset it represents, 
and the ideal application it provides to problems of parameter estimation. 
A welcome by-product of this approach is the ease with which the $U(1)'$ coupling can be varied, the interference effects can be included, and the exclusion limits for anticipated future runs of various collider experiments can be predicted.  

We are looking forward to carry out similar analyses with the first results by the CMS~\cite{CMS:2011wq} and ATLAS~\cite{Collaboration:2011xp} Collaborations.
There will also be further results (see, \eg Refs.~\cite{Abazov:2010ti,Aaltonen:2011gp}) 
and different channels by the CDF and D\O\ Collaborations.
Finally, a more systematic classification and discussion of the model class of Section~\ref{sec:models} is also underway~\cite{ER:2011}. 

\section*{Acknowledgments} 
The work at IF-UNAM is supported by CONACyT project 82291--F. 
The work of P.L. is supported by an IBM Einstein Fellowship and by NSF grant PHY--0969448.
E.R. acknowledges financial support provided by DGAPA--UNAM.

\appendix

\section{Cross-sections and PDFs}
\label{app:formalism}

The NLO differential cross-section for the DY process with a neutral gauge boson $G$ as the mediator,
$pp\rightarrow GX\rightarrow l^+ l^- X$, is given as~\cite{Ellis:1991qj},
\bea
\label{eq:nlo}
\frac{d\sigma^{\sbx{0.6}{NLO}}}{dM\ \ \ }
&=&\frac{2 }{N_c s}M\int dzdx_{1}\frac{1}{x_1z}\theta\left(1-\frac{1}{x_{1}zr^{2}_z}
\right)
\sum_{q} \hat{\sigma}_{q\bar{q} \to \ell^+\ell^-}(M^2)\\\nn
&\times&\Big[\Big\{f_{q}^{A}(x_{1},M^2)f_{\bar{q}}^{B}(x_{2},M^2)+f_{\bar{q}}^{A}(x_{1},M^2)f_{q}^{B}(x_{2},M^2)\Big\}\\\nn
&\times & \Big\{\delta(1-z)+\frac{\alpha_{s}(M^2)}{2\pi}D_{q}(z)\Big\}\\\nn
&+&\Big\{f_{g}^{A}(x_{1},M^2)[f_{q}^{B}(x_{2},M^2)+f_{\bar{q}}^{B}(x_{2},M^2)]\\\nn
&+&\ f_{g}^{B}(x_{2},M^2)[f_{q}^{A}(x_{1},M^2)+f_{\bar{q}}^{A}(x_{1},M^2)]\Big\}
\times \frac{\alpha_{s}(M^2)}{2\pi}D_{g}(z)\Big],
\eea
where $N_c=3$ is the color factor, $M$ is the invariant mass of the observed lepton pair and $\sqrt{s}$ is the energy of 
the $p\bar{p}$ collision in the CM frame, $r_z \equiv \sqrt{s}/M$, and $x_2^{-1} \equiv x_1z r_z ^2$. 
$f_{q/g}^A$ are the PDFs of the quarks and gluons coming from hadron $A$. 
$\alpha_s$ is the strong coupling constant, and
\bea
D_{q}(z)&=&C_{F}\Big[4(1+z^{2})\Big\{\frac{\log(1-z)}{1-z}\Big\}_{+}-2\frac{1+z^{2}}{1-z}\log z
         +\delta(1-z)\Big\{\frac{2\pi^2}{3}-8\Big\}\Big],\\\nn
D_{g}(z)&=&T_{R}\Big[\Big\{z^2+(1-z)^2\Big\}\log\frac{(1-z)^2}{z}+\frac{1}{2}+3z-\frac{7}{2}z^2\Big], 
\eea
with the $C_F=4/3$ and $T_R=1/2$, and the `+' distribution defined as
\be
\int_{0}^{1}dz g(z)\Big\{\frac{\log(1-z)}{1-z}\Big\}_{+}\equiv\int_{0}^{1}dz \Big\{g(z)-g(1)\Big\}\Big\{\frac{\log(1-z)}{1-z}\Big\}.
\ee
The expression for the hard scattering cross-section of the process $q\bar{q} \to \ell^+ \ell^-$,
\be
\label{eq:hard}
\hat{\sigma}_{q\bar{q} \to \ell^+\ell^-} (M^2)= \int_{-1}^{1}\frac{d\hat{\sigma}}{d\cos \theta^*}d\cos \theta^* 
\ee
$$
= \int_{-1}^{1} {d\cos \theta^*\over 128\pi M^2}\Big\{
\left(\lvert A_{LL}\rvert^2+\lvert A_{RR}\rvert^2\right)(1+\cos\theta^*)^2
+\left(\lvert A_{LR}\rvert^2+\lvert A_{RL}\rvert^2\right)(1-\cos\theta^*)^2\Big\},
$$
is given in terms of the polar angle, $\theta^*$, in the CM frame and the individual amplitudes,
\be
\label{eq:amplitud}
   A_{ij} = - Q(q) e^2 
             + {g_Z^2\, \epsilon_i(q) \epsilon_j(\ell) M^2 \over M^2 - M_Z^2 + i M_Z \Gamma_Z}
             + {g'^2\, Q'_i(q) Q'_j(\ell) M^2 \over M^2 - M_{Z'}^2 + i M_{Z'} \Gamma_{Z'}},
\ee
where $i,j$ run over $L,R$. 
$Q(q)$ is the electric charge of the quark and $e = g \sin\theta_W$. 
$M_{Z,Z'}$ and $\Gamma_{Z,Z'}$ are the masses and total decay widths of the $Z$ and $Z'$ bosons. 
\be 
   \epsilon_L (f) = T_3 (f) - Q(f)\sin^2\theta_W,   \hspace{24pt}  
   \epsilon_R (f) = - Q(f)\sin^2\theta_W,
\ee
are the effective couplings of the ordinary $Z$ to fermion $f$ entering with coupling strength, $g_Z = g/\cos\theta_W =  0.7433$.
The $Q'_L(f)$ are given in Table~\ref{tab:SMcharges} and $Q'_R(f) = -Q'_L(\bar{f})$.
As for the $Z'$ coupling strength, one often employs the (one-loop) unification value~\cite{Robinett:1982tq},
$g' = \sqrt{5/3}\sin\theta_W g_Z = 0.4615$, but we will do so only when comparing results to other analyses,
since one of our goals is to study the $g'$ dependence.

Integrating Eq.~(\ref{eq:nlo}) gives the LO differential 
cross-section\footnote{The integration over $z$ is carried out as $\int_0^{(1+\epsilon)} dz \delta(1-z)=1.$} as
\begin{align}
\label{eq:lo4}
\frac{d\sigma^{\sbx{0.6}{LO}}}{dM\ \ } =&\frac{2}{N_cs}M\int_{1/{r_z^2}}^1 dx_{1}\notag\\
\times&\frac{1}{x_1}
\sum_{q} \hat{\sigma}(M^2)\Big\{f_{q}^{A}(x_{1},M^2)f_{\bar{q}}^{B}(x_{2},M^2)+f_{\bar{q}}^{A}(x_{1},M^2)f_{q}^{B}(x_{2},M^2)\Big\},
\end{align}
where $x_{2}=\frac{1}{x_1r_z^2}$ now. 
Note here that for purely SM contribution to the process, the third term, while for the expected cross-section via a $Z'$ 
boson only, the first and second terms in Eq.~(\ref{eq:amplitud}) can simply be ignored. 
The decay width, $\Gamma_{Z'}$, given in eq.~(\ref{eq:amplitud}), is the sum of the partial decay widths of the $Z'$ boson into all the 
fermions it couples to. The partial decay width into a Dirac fermion pair is written as~\cite{Kang:2004bz}
\bea
\Gamma_{Z'\rightarrow  f \bar{f}}(M^2) &=& {g'^{2} M_{Z'} \over {24\pi}} \sqrt{1 - {4 M_f^2 \over M_{Z'}^2}}
\left[ \left( 1-{M_f^2\over M_{Z'}^2} \right)({Q'}_l^2 + {Q'}_r^2) + {6 M_f^2 \over M^2}Q'_l Q'_r \right] \frac{M^2}{M_{Z'}^2}, \hspace{24pt}
\label{eq:ZffD}
\eea
where $M_f$ is the mass of the final-state fermion.
We add the factor $M^2/M_{Z'}^2$ to get an `$\hat{s}$'-dependent $Z^{\prime}$-width~\cite{Baur:2001ze}.
For the range of $M_{Z'}$ of interest here, $M_f \ll M_{Z'}$ for SM fermions, and the 
above expression becomes independent of the fermion masses. 

With the above definitions we always obtain an effective NNLO result 
by multiplying by the corresponding $K^{m/n}$ factor in Eq.~(\ref{kfactor}). 
For NLO and LO PDFs the corresponding effective NNLO  cross-sections are, respectively,
\begin{align}
\label{eq:lo3}
\sigma_{Z'}^{\sbx{0.6}{NLO-PDF}}= \int_{M_{\sbx{0.6}{min}}}^{\sqrt{s}}dM K^{2/1}\frac{d\sigma^{\sbx{0.6}{NLO}}}{dM\ \ },
\hspace{50pt}
\sigma_{Z'}^{\sbx{0.6}{LO-PDF}}= \int_{M_{\sbx{0.6}{min}}}^{\sqrt{s}}dM K^{2/0}\frac{d\sigma^{\sbx{0.6}{LO}}}{dM\ \ },
\end{align}
where $M_{\sbx{0.6}{min}} = 0.1$~TeV is the lower  invariant mass of the analyzed events in~\cite{Aaltonen:2008ah}. 

\bibliographystyle{apsrev4-1}
\bibliography{myreferences2}

\end{document}